\documentclass[11pt,a4paper]{article}

\usepackage[a4paper,margin=2.5cm]{geometry} 

\usepackage[utf8]{inputenc} 
\usepackage[T1]{fontenc}
\usepackage{palatino} 
\usepackage{amsmath,amssymb,amsfonts,bm,latexsym}
\usepackage{graphicx}
\usepackage{subcaption}
\usepackage{textcomp}
\usepackage{float}
\usepackage{booktabs}
\usepackage{mathtools}
\usepackage{dcolumn}
\usepackage{ragged2e}
\usepackage{xcolor}
\usepackage{hyperref}
\usepackage{authblk} 
\usepackage{cite}

\DeclareUnicodeCharacter{03B1}{$\alpha$}

\hypersetup{
    colorlinks=true,
    linkcolor=red,
    citecolor=blue,
    filecolor=magenta,      
    urlcolor=cyan,
    pdftitle={Dynamical dark energy in models close to LCDM},
    pdfpagemode=FullScreen
}

\title{\textbf{Dynamical dark energy in models with evolution close to $\Lambda$CDM}}

\author[1,2]{Saikat Chakraborty\thanks{Corresponding author: saikat.c@chula.ac.th; saikat.chakraborty@nwu.ac.za}}
\author[3]{Charlotte Louw\thanks{lwxcha023@myuct.ac.za}}
\author[3,2,4]{Peter K.S. Dunsby\thanks{peter.dunsby@uct.ac.za}}
\author[3]{Kelly MacDevette\thanks{MCDKEL004@myuct.ac.za}}
\author[3,5]{Alvaro de la Cruz Dombriz\thanks{alvaro.delacruzdombriz@uct.ac.za}}

\affil[1]{High Energy Physics Theory Group, Department of Physics, Faculty of Science, Chulalongkorn University, Bangkok 10330, Thailand}
\affil[2]{Centre for Space Research, North-West University, Potchefstroom 2520, South Africa}
\affil[3]{Department of Mathematics and Applied Mathematics, Cosmology and Gravity Group, University of Cape Town, Rondebosch 7701, Cape Town, South Africa}
\affil[4]{South African Astronomical Observatory, Observatory 7925, Cape Town, South Africa}
\affil[5]{Departamento de F\'isica Fundamental, Universidad de Salamanca, 37008 Salamanca, Spain}

\date{} 

\begin{document}
\maketitle

\begin{abstract}
In this communication we address whether or not there is an equivalence between the kinematical and dynamical descriptions 
of the spatially flat $\Lambda$CDM model. We address this by investigating whether an almost $\Lambda$CDM expansion history ($j(z)\approx1$) corresponds to an almost $\Lambda$CDM model ($w_{\rm DE}(z)\approx-1$) by considering two particular explicit examples. At least for the cases considered, this turns out not to be the case. Instead, what we find is that an almost $\Lambda$CDM cosmic evolution rather corresponds to an \emph{almost unified dark fluid model}. Considering that one never gets the exact condition $j(z)=1$ from any cosmographic data sets, this raises further questions on whether the $\Lambda$CDM model is the best candidate for the standard model of the evolution of the universe.
\end{abstract}

\section{Introduction}

The nature of dark energy remains one of the most intriguing problems in modern cosmology. Observations from Type Ia supernovae \cite{Riess1998, Perlmutter1999} first established that in order for a Friedmann-Robertson-Walker (FRW) geometry to be compatible with the data, the universe is accelerating in the current epoch. This prompted the standard interpretation of dark energy as a cosmological constant $\Lambda$ within the framework of the $\Lambda$CDM or Concordance cosmological model. However, despite its observational success, the $\Lambda$CDM model faces theoretical challenges, particularly the fine-tuning and coincidence problems \cite{Weinberg1989, Martin2012}. As a result, numerous alternative models involving dynamical dark energy have been proposed to explain cosmic acceleration without invoking a strictly constant $\Lambda$.

Dynamical dark energy models often introduce scalar fields to drive cosmic acceleration. Quintessence \cite{Ratra1988} describes a slowly rolling 
scalar field with a potential $V(\phi)$ that allows for a time-dependent equation of state $w = p/\rho$. Unlike a cosmological constant with $w = -1$, quintessence models typically yield $-1< w <-1/3$ and can evolve over cosmic time. Some potentials, such as exponential or 
inverse power-law forms, offer tracking solutions that reduce fine-tuning concerns \cite{Zlatev1999}.

Tracker models \cite{Steinhardt1999} represent a subclass of quintessence models in which the scalar field dynamics naturally evolve to mimic the 
dominant energy component at early times before transitioning to dark energy domination. These models exhibit attractor-like behavior, making their 
late-time evolution largely insensitive to initial conditions. The advantage of tracker fields lies in their ability to alleviate the coincidence 
problem by allowing dark energy to evolve alongside matter, rather than emerging abruptly at late times.

Phantom dark energy \cite{Caldwell2002} represents another scalar field scenario in which $w < -1$, requiring a field with a negative kinetic term. Such models predict a future ``big rip’’ singularity where the expansion accelerates indefinitely \cite{Caldwell2003}. More general k-essence 
models \cite{ArmendarizPicon2000} extend scalar field dynamics by introducing non-canonical kinetic terms, allowing for varying sound speeds 
and alleviating some fine-tuning issues.

Dynamical dark energy can also emerge from modifications to General Relativity (GR). One well-studied class of models involves $f(R)$ gravity \cite{Carroll2004, Starobinsky2007}, where the Ricci scalar $R$ in the Einstein-Hilbert action is replaced by a function $f(R)$, leading to an 
effective dark energy component. Similarly, scalar-tensor theories such as Brans-Dicke gravity \cite{Brans1961} introduce a varying gravitational 
coupling, which can mimic dynamical dark energy behavior \cite{Fujii2003}.

Other approaches include higher-dimensional theories such as DGP gravity \cite{Dvali2000}, which conjectures that our 4D universe is embedded in a higher-dimensional spacetime. In this model, the leakage of gravity into extra dimensions at large scales can drive late-time acceleration. Massive 
gravity \cite{deRham2011} and Horndeski theories \cite{Horndeski1974, DeFelice2010} also provide frameworks where modified gravity gives rise to 
dynamical dark energy effects.

Dynamical dark energy models are subject to stringent observational tests. Cosmic microwave background (CMB) anisotropies \cite{Planck2020}, baryon acoustic oscillations \cite{Eisenstein2005}, and large-scale structure surveys \cite{DES2018} provide constraints on the equation of state parameter $w$ and its evolution. Current data generally favor a nearly constant $w\approx -1$, but DESI, Euclid, the Vera C. Rubin Observatory and 
the Nancy Grace Roman Space Telescope will refine constraints on dynamical dark energy evolution. Indeed, the first and second year data release of DESI appears to 
favor evolution in the dark energy equation of state \cite{DESI:2025zgx}. 

Cosmography, an approach based on a model-independent expansion of cosmological observables, provides further constraints on dynamical dark energy by directly analyzing the kinematics of cosmic expansion without assuming a specific gravitational theory \cite{Dunsby:2015ers}. By parameterizing the Hubble expansion in terms of series coefficients such as the deceleration q, jerk j, and snap s parameters \cite{Visser2004}, cosmographic methods allow for a direct test of 
deviations from $\Lambda$CDM. Recent studies \cite{Capozziello2019,Aviles2020,Busti:2015xqa,delaCruz-Dombriz:2016bqh} have explored how these higher-order parameters can 
differentiate between dark energy models, offering additional insights into the nature of cosmic acceleration.

Any cosmological model in General Relativity can be specified as an algebraic cosmographic constraint \cite{Dunajski:2008tg,Chakraborty:2021jku,Chakraborty:2021mcf}. For example, 
$\Lambda$CDM model, in its full generality, is characterized by
    \begin{equation}
        s + 2(q+j) + qj = 0\,.
    \end{equation}
The basic idea is to reinterpret the parameters appearing in the cosmological field equations as integration constants and specify a cosmological solution in a parameter-independent way involving only cosmographic quantities. Since we can constrain the present-day values of the cosmographic parameters from the Luminosity distance-Redshift data in a model-independent manner, this cosmographic method of specifying a cosmological solution is appealing. In particular, the spatially flat $\Lambda$CDM model is dynamically characterized by spatial flatness and a dark energy equation of state $w_{\rm DE}(z)=-1$. On the other hand, kinematically it is characterized by the cosmographic condition $j(z)=1$. The equivalence between the kinematical and the dynamical description is the basis of statefinder diagnostic \cite{Sahni:2002fz,Sahni:2002yq,Alam:2003sc}, which attributes the condition $j(z)=1$ to $\Lambda$CDM model. 

However, something that is often overlooked in the literature is that the kinematical and dynamical descriptions are not guaranteed to be equivalent. In particular, there is not a one-to-one correspondence between the kinematical condition $j(z)=1$ and the dynamical condition $w_{\rm DE}(z)=-1$. There exists a distinct class of dynamical dark energy models called the unified dark fluid model, which are kinematically completely equivalent to $\Lambda$CDM (i.e. compatible with the cosmographic condition $j(z)=1$). This is a barotropic fluid model with constant pressure and vanishing sound speed \cite{Luongo:2014nld}, whose equation of state parameter is inversely proportional to its energy density. The energy density of such a fluid behaves as pressureless matter in the early time and as an emerging cosmological constant in the asymptotic future, thus providing a unified framework for dark matter and dark energy. In a sense, this is equivalent to (generalized) Chaplygin gas models, but interestingly they are kinematically degenrate with $\Lambda$CDM. We show this degeneracy explicitly in this paper, showing that the condition $j(z)=1$ is equivalent to the condition $w_{\rm DE}(z)=-1$ only if one assumes $w_{\rm DE}(0)=-1$.
    
The purpose of this articles is therefore to question the robustness of this very equivalence between the kinematical and the dynamical description of the spatially flat $\Lambda$CDM model. We stated in the last paragraph that they are equivalent given one assumes $w_{\rm DE}(0)=-1$. The question we address here is: 
\begin{itemize}
    \item \emph{Given that an actual $\Lambda$CDM evolution ($j(z)=1$) corresponds to the actual $\Lambda$CDM model ($w_{\rm DE}(z)=-1$), does an almost $\Lambda$CDM evolution ($j(z)\approx1$) correspond to an almost $\Lambda$CDM model ($w_{\rm DE}(z)\approx-1$)?}
\end{itemize}
Considering two particular explicit examples, we find the answer to be non-affirmative. What we find is that an almost $\Lambda$CDM cosmic evolution rather corresponds to an \emph{almost unified dark fluid model}. 

Given that we never actually get exactly $j(z)=1$, but rather get $j(z)\approx1$ while trying to obtain a fitting function from data sets like BAO, CMB, CC, Pantheon etc \cite{Mukherjee:2020ytg}, our analysis shows that it would be na\'ive for us to conclude, based on these data sets, that the underlying dark energy \emph{model} is something close to the $\Lambda$CDM model. In general, our finding indicates that the equivalence of the kinematic description vis-a-vis cosmography and the dynamical description vis-a-vis the dark energy equation of state is not robust against small uncertainties in the astrophysical measurements of the cosmographic parameters. This puts a question mark on the validity of the practice of specifying a model itself based on its cosmography, e.g. the idea which the statefinder diagnostic is based upon \cite{Sahni:2002fz,Sahni:2002yq,Alam:2003sc}. 

The article is organized as follows. Section \ref{sec:kin_vs_dyn} starts with a discussion on the (in)equi\-valence of the kinematical cosmographic condition $j(z)=1$ and the dynamical condition $w_{\rm DE}(z)=-1$. In section \ref{sec:almost_LCDM} we introduce the two explicit examples of almost $\Lambda$CDM-like evolutionary models that we consider in this paper, and calculate various dynamical quantities of interest corresponding to them. We present the plots of these dynamical quantities in section \ref{sec:cosm_evoln} to explicitly show their kinematic closeness to the exact $\Lambda$CDM-like cosmic evolution, and distinct dynamics from the exact $\Lambda$CDM model. Section \ref{sec:DESI} is devoted to confronting the two almost $\Lambda$CDM evolutionary models with the latest DESI data release. We conclude in section \ref{sec:conclusion}, summarizing our findings and discussing possible implications in cosmological model building.

\section{On the conditions $j(z)=1$ and $w_{\rm DE}(z)=-1$}\label{sec:kin_vs_dyn}
The cosmological field equations for a flat $\Lambda$CDM universe model can be written as 
\begin{subequations}
    \begin{eqnarray}\label{field_eqs}
        3H^{2} = \rho + \rho_{\rm DE}\;,\label{eq:fried}
        \\
        -\left(2\dot{H} + 3H^{2}\right) = P_{\rm DE}\,,\label{eq:Raychoudhuri_new}
    \end{eqnarray}     
\end{subequations}
where $\rho$ is the energy density of non-relativistic matter, $\rho_{\rm DE}$ \& $P_{\rm DE}$ are the energy density and pressure of Dark Energy, and the dot means derivative with respect to cosmic time. 
Using the above field equations one can write
\begin{equation}\label{DE_eos}
    w_{\rm DE} = \frac{P_{\rm DE}}{\rho_{\rm DE}} = - \frac{2\dot{H}+3H^{2}}{3H^{2}-\rho} = \frac{H^{2}-\frac{R}{3}}{3H^{2}-\rho}\,.
\end{equation}
In this form, the non-relativistic matter and the dark energy is separately conserved
\begin{subequations}\label{eq:cons}
    \begin{eqnarray}
        && \dot\rho + 3H\rho = 0\,.\\
        && \dot\rho_{\rm DE} + 3H\rho_{\rm DE}(1+w_{\rm DE}) = 0\,.
    \end{eqnarray}
\end{subequations}
Dividing the numerator and denominator by $H^2$ we arrive at
\begin{equation}\label{eq:wDE}
 w_{\rm DE} = \frac{2q-1}{3-3\Omega_m}\,,  
\end{equation}
where in the last step we have utilized the relation
\begin{equation}
    \frac{R}{6H^2} = 1-q\,.
\end{equation}
and $\Omega_m\equiv\frac{\rho}{3H^2}$ is the usual matter density abundance parameter. $\Omega_m$ obeys the dynamical equation
\begin{equation}\label{eq:dyn_Omegatilde}
    \frac{{\rm d}\Omega_m}{{\rm d}\tau} = -\Omega_m (1-2q)\,.
\end{equation}
where ${\rm d}\tau=H{\rm d}t$. Now, imposing $w_{\rm DE}=-1$ one obtains
\begin{equation}
    2q - 3\Omega_m=-2\,.
\end{equation}
Taking a $\tau$-derivative, utilizing \eqref{eq:dyn_Omegatilde} and using the relation
\begin{equation}
    \frac{{\rm d}q}{{\rm d}\tau} = 2q^2 +q - j\,,
\end{equation}
we can arrive at the condition $j=1$, after performing some rather straightforward steps. This remarkable condition exactly characterizes the 
dynamics of a flat $\Lambda$CDM model. 

Consider, now, a dynamical dark energy model with $w_{\rm}=-\alpha(\tau)$. Imposing this in \eqref{eq:wDE} and taking a $\tau$-derivative, we obtain
\begin{equation}
    j-1 = \left(q-\frac{1}{2}\right)\left(3-3\alpha-\frac{{\rm d}\alpha/{\rm d}\tau}{\alpha}\right)\,.
\end{equation}
One possibility to get $j=1$, i.e., kinematic degeneracy with $\Lambda$CDM is $q=\frac{1}{2}$, but this cannot be true for the entire evolution 
history. Another possibility is $\alpha=1$, which takes us back to the case $w_{\rm DE}=-1$. The third possibility is that 
$\alpha(\tau)$ satisfies the equation
\begin{equation}
    \frac{{\rm d}\alpha}{{\rm d}\tau} = 3\alpha(1-\alpha)\,,
\end{equation}
which, upon solving, gives 
\begin{equation}
    \alpha(\tau) = \frac{\alpha_0 {\rm e}^{3\tau}}{\alpha _0 e^{3\tau} + (1-\alpha_0)}\,,
\end{equation}
with $\alpha_0=\alpha(0)$. This implies a dark energy equation of state that can be expressed as 
\begin{equation}\label{eq:DDE}
    w_{\rm DE}(\tau) = -\frac{w_{\rm DE}(0){\rm e}^{3\tau}}{w_{\rm DE}(0){\rm e}^{3\tau} - (1+w_{\rm DE}(0))} \Leftrightarrow w_{\rm DE}(z) = -\frac{1}{1 - \left(\frac{1+w_{\rm DE}(0)}{w_{\rm DE}(0)}\right)(1+z)^3}\,.
\end{equation}
What Eq.\eqref{eq:DDE} implies is that generically the cosmographic condition $j=1$ corresponds to a situation where the additional component, apart from the non-relativistic matter, actually behaves as a mixture of dark matter and dark energy. One can go one step further and calculate from the dark energy conservation equation of \eqref{eq:cons} that
\begin{equation}
     \rho_{\rm DE} \propto \frac{1}{\alpha} = -\frac{1}{w_{\rm DE}} = -\left(\frac{1+w_{\rm DE}(0)}{w_{\rm DE}(0)}\right)
     \frac{1}{a^3}+ 1\,, \qquad P_{\rm DE}=constant\,.
\end{equation}
i.e. it corresponds to a dynamical dark energy model in which the dark energy equation of state varies inversely with its energy density, so that its pressure remains constant. This is the so-called unified dark fluid model \cite{Luongo:2014nld} with vanishing sound speed. One can notice that the energy density of the fluid scales partially as $~(1+z)^0$ and partially as $~(1+z)^3$, enabling it to behave as a pressureless matter in the asymptotic past and as a cosmological constant in the asymptotic future. Therefore, the fluid can explain, in a unified manner, both dark matter and dark energy, justifying its name as the unified dark fluid.\footnote{The $\Lambda$CDM-model and the unified dark fluid model both are freezing models ($w_{\rm DE}\to-1$) with $\rho_{\rm DE}\to\rho_0$(constant) asymptotically, while conforming to the cosmographic condition $j=1$. If the latter cosmographic condition is relaxed, then freezing models can also have $\rho_{\rm DE}\to0$ asymptotically \cite{Scherrer:2018cdx}.}

If one sets $w_{\rm DE}(0)=-1$ in Eq.\eqref{eq:DDE}, then one gets $w_{\rm DE}(z)=-1$. According to Eq.\eqref{eq:wDE}, setting $w_{\rm DE}(0)=-1$ implies setting $\Omega_{m0}=\frac{2}{3}(1+q_0)$. The cosmographic condition $j(z)=1$ exactly corresponds to the $\Lambda$CDM \emph{model} itself only when $\Omega_{m0}=\frac{2}{3}(1+q_0)$. Otherwise, generically, it corresponds to a dark fluid model.

The question we would like to address is the following: 
\begin{itemize}
    \item \emph{Even if an exact $\Lambda$CDM-like cosmic evolution ($j(z)=1$) correspond to the exact $\Lambda$CDM model ($w_{\rm DE}(z)=-1$), does an almost $\Lambda$CDM-like cosmic evolution ($j(z)\approx1$) imply an almost $\Lambda$CDM model ($w_{\rm DE}\approx-1$)?}
\end{itemize}
If we call the cosmographic condition $j(z)=1$ the kinematical specification of the $\Lambda$CDM model and the condition $w_{\rm DE}(z)=-1$ the dynamical specification of the $\Lambda$CDM model, then the question can be framed more formally:
\begin{itemize}
    \item \emph{Does kinematical proximity to the $\Lambda$CDM model translate to dynamical closeness to the $\Lambda$CDM model?} 
\end{itemize}
Na\'ively one may expect an affirmative answer. However, as we will show, using two explicit examples, this is not the case.

\section{Almost $\Lambda$CDM cosmologies}\label{sec:almost_LCDM}

The most general solution to the cosmographic condition $j(z)=1$ can be expressed in terms of the Hubble parameter as \cite{Zhai:2013fxa}
\begin{equation}\label{lcdm_H}
    H^2(z) = H_0^2 \left[c_1(1+z)^3 + (1-c_1))\right]\,,
\end{equation}
Or, in terms of the scale factor as \cite{Choudhury:2019zod}
\begin{equation}\label{lcdm_a}
    a(t) = a_0\left[\cosh (\lambda t) + c \sinh (\lambda t)\right]^{\frac{2}{3}}\,.
\end{equation}
The deceleration parameter is given by
\begin{equation}\label{dec_param}
q(t) = -1-\frac{\dot{H}(t)}{H(t)^2} \quad \text{or} \quad q(z)=-1+(1+z)\frac{h'(z)}{h(z)}\,,
\end{equation}
where prime means derivative with respect to redshift. 
If we want the present-day value of the deceleration parameter to be $q(z=0)=q_0$, then the constant $c_1$ in Eq.\eqref{lcdm_H} comes out to be $c_1=\frac{2}{3}(1+q_0)$. 

Based on the above, we consider two prototype examples for an \emph{almost $\Lambda$CDM} evolutionary models
\begin{enumerate}
    \item Evolutionary model-I: 
    \begin{equation}\label{almostlcdm_1}
        a(t) = a_0\left[\cosh (\lambda t) + c \sinh (\lambda t)\right]^{\frac{2}{3(1+\epsilon)}}\,.
    \end{equation}
    \item Evolutionary model-II:
    \begin{equation}\label{almostlcdm_2}
        H^2(z) = H_0^2 \left[c_1(1+z)^{\frac{1}{2} \left(3 + \sqrt{8 \epsilon +9}\right)} + (1-c_1)(z+1)^{\frac{1}{2} \left(3-\sqrt{8 \epsilon +9}\right)}\right]\,.
    \end{equation}
\end{enumerate}

Jerk parameter is given by
\begin{equation}\label{jerk_param}
    j(t) = 2q(t)^2 + q(t) - \frac{\dot{q}(t)}{H(t)} \quad \text{or} \quad j(z) = 2q(z)^2 + q(z) + (1+z)q'(z)\,,
\end{equation}
where prime means derivative with respect to redshift. Using Eqs.\eqref{dec_param} and \eqref{jerk_param} one can compute the corresponding cosmographic conditions:
\begin{enumerate}
    \item Evolutionary model-I: 
    \begin{equation}\label{cosm_cond_1}
        j=1+3\epsilon(1+q)\,.
    \end{equation}
    \item Evolutionary model-II: 
    \begin{equation}\label{cosm_cond_2}
        j=1+\epsilon\,.
    \end{equation}
\end{enumerate}

\subsection{Evolutionary model-I: $a(t) = a_0(\cosh (\lambda  t) + c\sinh (\lambda  t))^{\frac{2}{3 (\epsilon +1)}}$}\label{subsec:example-I}

For this example, we start from the cosmographic condition \eqref{cosm_cond_1}.
We solve the equation
\begin{equation}
(1+z)\frac{{\rm d}q}{{\rm d}z} = - 2q^2 - q + j = - 2q^2 - q + 1 + 3\epsilon(1+q)
\end{equation}
to obtain
\begin{equation}
q(z) = \frac{(3 \epsilon +1) (z+1)^{3 \epsilon +3}-{\rm e}^{3 c_1 (\epsilon +1)}}{{\rm e}^{3 c_1 (\epsilon +1)}+2 (z+1)^{3 \epsilon +3}}\,.
\end{equation}
Next, we solve the equation
\begin{equation}
    (1+z)\frac{dh}{dz} = h (q+1)\,,
\end{equation}
where the integration constant has been fixed such that $h(z=0)=1$.
\begin{equation}
    h(z) = \frac{\sqrt{{\rm e}^{3 c_1 (\epsilon +1)}+2 (z+1)^{3 \epsilon +3}}}{\sqrt{{\rm e}^{3 c_1 (\epsilon +1)}+2}}\,.
\end{equation}
From Eq.\eqref{almostlcdm_1} we obtain
\begin{equation}
    j(z) = \frac{9 \epsilon  (\epsilon +1) (z+1)^{3 \epsilon +3}}{{\rm e}^{3 c_1 (\epsilon +1)}+2 (z+1)^{3 \epsilon +3}}+1\,.
\end{equation}
Solving the equation
\begin{equation}
    (1+z)\frac{{\rm d}\Omega_m}{{\rm d}z} = (1-2q)\Omega _m
\end{equation}
we get
\begin{equation}
 \Omega_m(z) = \frac{c_2 (z+1)^3}{{\rm e}^{3 c_1 (\epsilon +1)}+2 (z+1)^{3 \epsilon +3}}\,.
\end{equation}
Using Eq.\eqref{eq:wDE} we can write
\begin{equation}
    w_{\rm DE}(z) = \frac{2 \epsilon  (z+1)^{3 \epsilon +3}-{\rm e}^{3 c_1 (\epsilon +1)}}{(z+1)^3 \left(2 (z+1)^{3 \epsilon }-c_2\right)+{\rm e}^{3 c_1 (\epsilon +1)}}\,.
\end{equation}
We note that in all the above expressions the constant $c_1$ appear only as the combination $e^{3c_1(1+\epsilon)}$. Redefining the constant as $e^{3c_1(1+\epsilon)}\to c_1$ makes the expressions look considerably simpler. Hence, we list the expressions below
\begin{subequations}
    \begin{eqnarray}
        h^2(z) &=& \frac{c_1 + 2(1+z)^{3 \epsilon +3}}{c_1+2}\,,
        \\
        q(z) &=& \frac{(3 \epsilon +1) (z+1)^{3 \epsilon +3}-c_1}{c_1+2 (z+1)^{3 \epsilon +3}}\,,
        \\
        j(z) &=& 3 \epsilon  \left[\frac{(3 \epsilon +1) (z+1)^{3 \epsilon +3}-c_1}{c_1+2 (z+1)^{3 \epsilon +3}}+1\right]+1\,,
        \\
        \Omega_m(z) &=& \frac{c_2 (z+1)^3}{c_1+2 (z+1)^{3 \epsilon +3}}\,.
        \\
        w_{\rm DE}(z) &=& \frac{2 \epsilon  (z+1)^{3 \epsilon +3}-c_1}{c_1+(z+1)^3 \left(2 (z+1)^{3 \epsilon }-c_2\right)}
    \end{eqnarray}
\end{subequations}
In total, there are three unknown parameters in the above expressions: two integration constants $c_1,c_2$ and the parameter $\epsilon$ that characterises deviation from the exact $\Lambda$CDM-like cosmic evolution. They can be related to the present-day value of the deceleration, jerk and the matter density abundance parameter by setting
\begin{equation}
    \{q(0),j(0),\Omega_{m}(0)\}=\{q_0,j_0,\Omega_{m0}\}\,.
\end{equation}
Solving these three equations simultaneously one gets
\begin{equation}\label{const_redef-I}
\epsilon = \frac{j_0-1}{3 (q_0+1)},\quad c_1 = \frac{j_0-2q_0^2-q_0}{(q_0+1)^2},\quad c_2 = \frac{\Omega_{m0}(j_0+3q_0+2)}{(q_0+1)^2}
\end{equation}
Substituting them back into the expressions we get the following
\begin{subequations}\label{cosmology-I}
    \begin{eqnarray}
        h^2(z) &=& \frac{2 (1 + z)^{(2 + j_0 + 3 q_0)/(1 + q_0)} + [j_0 - q_0 (1 + 2 q_0)]/(1 + q_0)^2}{(2 + j_0 + 3 q_0)/(1 + q_0)^2}\,,
        \\
        q(z) &=& \frac{(q_0+1) (j_0+q_0) (z+1)^{\frac{j_0+3 q_0+2}{q_0+1}}-j_0+2 q_0^2+q_0}{(q_0+1)^2 \left[2 (z+1)^{\frac{j_0+3 q_0+2}{q_0+1}}+\frac{j_0-q_0 (2 q_0+1)}{(q_0+1)^2}\right]}\,,
        \\
        j(z) &=& \frac{(j_0-1) (j_0+3 q_0+2) (z+1)^{\frac{j_0+3 q_0+2}{q_0+1}}}{2 (q_0+1)^2 (z+1)^{\frac{j_0+3 q_0+2}{q_0+1}}+j_0-q_0 (2 q_0+1)}+1\,,
        \\
        \Omega_m(z) &=& \frac{\Omega_{m0} (z+1)^3 (j_0+3 q_0+2)}{(q_0+1)^2 \left[2 (z+1)^{\frac{j_0+3 q_0+2}{q_0+1}}+\frac{j_0-q_0 (2 q_0+1)}{(q_0+1)^2}\right]}\,,
        \\
        w_{\rm DE}(z) &=& \frac{(q_0+1)^2 \left[\frac{-j_0+2 q_0^2+q_0}{(q_0+1)^2}+\frac{2 (j_0-1) (z+1)^{\frac{j_0+3 q_0+2}{q_0+1}}}{3 (q_0+1)}\right]}{(z+1)^3 \left[2 (q_0+1)^2 (z+1)^{\frac{j_0-1}{q_0+1}}-\Omega_{m0} (j_0+3 q_0+2)\right]+j_0-2 q_0^2-q_0}\,.\label{wDE_cosmology-I}
    \end{eqnarray}
\end{subequations}
Although the above expressions look complicated, it is easier to verify that
\begin{equation}
    \{h(0),q(0),j(0),\Omega_m(0)\}=\{1,q_0,j_0,\Omega_{m0}\}\,.
\end{equation}
One can calculate the transition redshift $z_{\rm tr}$, at which the universe transitions from deceleration to acceleration, by solving $q(z_{\rm tr})=0$. In terms of the parameters $\{q_0,j_0,\Omega_{m0}\}$, $z_{\rm tr}$ is
\begin{equation}\label{ztr-I}
    z_{\rm tr} = \left[\frac{j_0-2 q_0^2-q_0}{(q_0+1) (j_0+q_0)}\right]^{\frac{q_0+1}{j_0+3 q_0+2}}-1
\end{equation}
One can also calculate the present day value of dark energy $w_{\rm DE}(0)$ and its derivative $w_{\rm DE}'(0)$ as
\begin{equation}\label{w0wa-I}
    w_{\rm DE}(0) = \frac{1-2 q_0}{3 (\Omega_{m0}-1)}\,,\qquad w_{\rm DE}'(0) = -\frac{2 j_0 (\Omega_{m0}-1)+2 q_0 (2 q_0-3 \Omega_{m0}+1)+\Omega_{m0}}{3 (\Omega_{m0}-1)^2}
\end{equation}

\subsection{Evolutionary model-II: $h^2(z) = c_1(1+z)^{\frac{1}{2} \left(3 + \sqrt{8 \epsilon +9}\right)} + (1-c_1)(z+1)^{\frac{1}{2} \left(3-\sqrt{8 \epsilon +9}\right)}$}\label{subsec:example-II}

For this example, we start with the expression of the Hubble parameter
\begin{equation}
    h^2(z) = c_1(1+z)^{\frac{3+\sqrt{9+8\epsilon}}{2}}+(1-c_1)(1+z)^{\frac{3-\sqrt{9+8\epsilon}}{2}}
\end{equation}
From \eqref{dec_param} one can calculate the deceleration parameter
\begin{equation}
    q(z)=\frac{-1-\sqrt{9+8\epsilon}+c_1\left[1-(1+z)^{\sqrt{9+8\epsilon}}+\sqrt{9+8\epsilon}+(1+z)^{\sqrt{9+8\epsilon}}\sqrt{9+8\epsilon}\right]}{4+4c_1\left[-1+(1+z)^{\sqrt{9+8\epsilon}}\right]}
\end{equation}
Solving the following differential equation 
\begin{equation}
    (1+z)\Omega_m'(z)=\Omega_m(z)(1-2q(z))\,,
\end{equation}
we get $\Omega_m$
\begin{equation}
\Omega_m(z)=\frac{c_2(1+z)^{\frac{\sqrt{9+8\epsilon}+3}{2}}}{1+c_1\left[-1+(1+z)^{\sqrt{9+8\epsilon}}\right]}\,.
\end{equation}
Using Eq.\eqref{eq:wDE} we can write
\begin{equation}
    w_{\rm DE}(z) = \frac{c_1 \left[\sqrt{8 \epsilon +9} (z+1)^{\sqrt{8 \epsilon +9}}-3 (z+1)^{\sqrt{8 \epsilon +9}}+\sqrt{8 \epsilon +9}+3\right]-\sqrt{8 \epsilon +9}-3}{6 c_1 \left[(z+1)^{\sqrt{8 \epsilon +9}}-1\right]-6 c_2 (z+1)^{\frac{1}{2} \left(\sqrt{8 \epsilon +9}+3\right)}+6}\,.
\end{equation}

Let us list below all the quantities that we have calculated
\begin{subequations}
    \begin{eqnarray}
    h^2(z) &=& c_1(1+z)^{\frac{3+\sqrt{9+8\epsilon}}{2}}+(1-c_1)(1+z)^{\frac{3-\sqrt{9+8\epsilon}}{2}}\,,
        \\
    q(z) &=& \frac{-1-\sqrt{9+8\epsilon}+c_1\left(1-(1+z)^{\sqrt{9+8\epsilon}}+\sqrt{9+8\epsilon}+(1+z)^{\sqrt{9+8\epsilon}}\sqrt{9+8\epsilon}\right)}{4+4c_1\left(-1+(1+z)^{\sqrt{9+8\epsilon}}\right)}\,,\nonumber
        \\
    &&    \\
    j(z) &=& 1+\epsilon\,,
        \\
   \Omega_m(z) &=& \frac{c_2(1+z)^{\frac{\sqrt{9+8\epsilon}+3}{2}}}{1+c_1\left[-1+(1+z)^{\sqrt{9+8\epsilon}}\right]}\,.
        \\
    w_{\rm DE}(z) &=& \frac{c_1 \left[\sqrt{8 \epsilon +9} (z+1)^{\sqrt{8 \epsilon +9}}-3 (z+1)^{\sqrt{8 \epsilon +9}}+\sqrt{8 \epsilon +9}+3\right]-\sqrt{8 \epsilon +9}-3}{6 c_1 \left[(z+1)^{\sqrt{8 \epsilon +9}}-1\right]-6 c_2 (z+1)^{\frac{1}{2} \left(\sqrt{8 \epsilon +9}+3\right)}+6}\,.\nonumber
    \\
    &&
    \end{eqnarray}
\end{subequations}
Again, there are three unknown parameters in the above expressions: two integration constants $c_1,c_2$ and the parameter $\epsilon$ that characterises deviation from the exact $\Lambda$CDM-like cosmic evolution. Setting
\begin{equation}
    \{q(0),j(0),\Omega_{m}(0)\}=\{q_0,j_0,\Omega_{m0}\}\,,
\end{equation}
they can be solved in terms of $\{q_0,j_0,\Omega_{m0}\}$ as
\begin{equation}\label{const_redef-II}
\epsilon = -1+j_0\,,\quad c_1 = \frac{4 \sqrt{8 j_0+1} q_0+8 j_0+\sqrt{8 j_0+1}+1}{2 (8 j_0+1)},\quad c_2 = \Omega_{m0}\,.
\end{equation}
Substituting them back into the expressions, we get the following
\begin{subequations}\label{cosmology-II}
    \begin{eqnarray}
        h^2(z) &=& (z+1)^{\frac{3}{2}-\frac{1}{2} \sqrt{8 j_0+1}} \left\{\frac{\left(\sqrt{8 j_0+1}+4 q_0+1\right) \left[(z+1)^{\sqrt{8 j_0+1}}-1\right]}{2 \sqrt{8 j_0+1}}+1\right\}\,,
        \\
        q(z) &=& \frac{\left(\sqrt{8 j_0+1}-1\right) q_0 (z+1)^{\sqrt{8 j_0+1}}+\left(\sqrt{8 j_0+1}+1\right) q_0+2 j_0 \left[(z+1)^{\sqrt{8 j_0+1}}-1\right]}{(4 q_0+1) (z+1)^{\sqrt{8 j_0+1}}+\sqrt{8 j_0+1} (z+1)^{\sqrt{8 j_0+1}}+\sqrt{8 j_0+1}-4 q_0-1}\,,\nonumber
        \\
        && \\
        j(z) &=& j_0\,,
        \\
        \Omega_m(z) &=& \frac{\Omega_{m0} (z+1)^{\frac{1}{2} \left(\sqrt{8 j_0+1}+3\right)}}{\frac{\left(\sqrt{8 j_0+1}+4 q_0+1\right) \left((z+1)^{\sqrt{8 j_0+1}}-1\right)}{2 \sqrt{8 j_0+1}}+1}\,,
        \\
        w_{\rm DE}(z) &=& \frac{\frac{\left(4 \sqrt{8 j_0+1} q_0+8 j_0+\sqrt{8 j_0+1}+1\right) \left(\sqrt{8 j_0+1} (z+1)^{\sqrt{8 j_0+1}}-3 (z+1)^{\sqrt{8 j_0+1}}+\sqrt{8 j_0+1}+3\right)}{16 j_0+2}-\sqrt{8 j_0+1}-3}{\frac{3 \left(\sqrt{8 j_0+1}+4 q_0+1\right) \left((z+1)^{\sqrt{8 j_0+1}}-1\right)}{\sqrt{8 j_0+1}}-6 \Omega_{m0} (z+1)^{\frac{1}{2} \left(\sqrt{8 j_0+1}+3\right)}+6}\,.\nonumber
        \\
        && \label{wDE_cosmology-II}
    \end{eqnarray}
\end{subequations}
Again, one can verify that
\begin{equation}
    \{h(0),q(0),j(0),\Omega_m(0)\}=\{1,q_0,j_0,\Omega_{m0}\}\,.
\end{equation}
As before, one can calculate the transition redshift $z_{\rm tr}$, at which $q$ changes sign from positive to negative, as
\begin{equation}\label{ztr-II}
    z_{\rm tr} = 2^{-\frac{1}{\sqrt{8 j_0+1}}} \left(\frac{2 j_0^2+4 j_0 q_0^2+\sqrt{8 j_0+1} q_0^2-2 \sqrt{8 j_0+1} j_0 q_0-2 j_0 q_0+q_0^2}{j_0^2-2 j_0 q_0^2-j_0 q_0}\right)^{\frac{1}{\sqrt{8 j_0+1}}}-1
\end{equation}
and the present-day value of dark energy $w_{\rm DE}(0)$ and its derivative $w_{\rm DE}'(0)$ as
\begin{equation}\label{w0wa-II}
    w_{\rm DE}(0) = \frac{1-2 q_0}{3 (\Omega_{m0}-1)}\,,\qquad w_{\rm DE}'(0) = -\frac{2 j_0 (\Omega_{m0}-1)+2 q_0 (2 q_0-3 \Omega_{m0}+1)+\Omega_{m0}}{3 (\Omega_{m0}-1)^2}
\end{equation}

\section{Cosmic evolution by specifying $q_0,j_0,\Omega_{m0}$}\label{sec:cosm_evoln}

For both the almost $\Lambda$CDM evolutionary models considered in the subsections \ref{subsec:example-I} and \ref{subsec:example-II}, we had three parameters to start with: two integration constants $c_1,c_2$ and one smallness parameter $\epsilon$ characterizing the deviation from the exact $\Lambda$CDM-like evolution ($j=1$). What we have been able to achieve is to replace the set $\{\epsilon,c_1,c_2\}$ with the new set $\{q_0,j_0,\Omega_{m0}\}$. 
From \eqref{cosmology-I} and \eqref{cosmology-II}, one can see that to show the evolution of various cosmological quantities graphically, one needs to first specify the present-day values of the deceleration, jerk and the matter abundance parameter $\{q_0,j_0,\Omega_{m0}\}$.  

By combining the Cosmic Chronometer data (\cite{Zhang:2012mp,Stern:2009ep,Moresco:2012jh,Moresco:2015cya,Moresco:2016mzx,Ratsimbazafy:2017vga}), Baryon Acoustic Oscillation data (\cite{Gaztanaga2009,Oka2014,Wang2017,Chuang2013,Alam2017,
Blake2012,Chuang2013b,Anderson2014,Zhao2019,Busca2013,Bautista2017,Delubac2015,FontRibera2014}), Pantheon SN1a compilation (\cite{Pan-STARRS1:2017jku}) and CMB data \cite{Planck:2015bue}, and employing a Gaussian process, one can obtain a model-independent estimate for $q_0,j_0$ typically as \cite[Table 5]{Mukherjee:2020ytg} 
\begin{equation}\label{q0j0_values}
    \{q_0,j_0\} = \{-0.647^{+0.070}_{-0.069},1.035^{+0.219}_{-0.227}\}\,.
\end{equation}
We take the following values of the parameters
\begin{equation}\label{param_values}
    \{q_0,j_0\} = \{-0.647,1.035\}\,, \quad \Omega_{m0} = \frac{2}{3}(1+q_0)\,,
\end{equation}
where the value of $\Omega_{m0}$ has been chosen such that $w_{\rm DE}(z=0)=-1$ for both the almost $\Lambda$CDM evolutionary models, as can be verified from Eqs. \eqref{wDE_cosmology-I} and \eqref{wDE_cosmology-II}. 

It can be calculated from Eqs.\eqref{const_redef-I} and \eqref{const_redef-II} that, for the chosen values of $\{q_0,j_0\}$, the values of the deviation parameter $\epsilon$ which quantifies deviation from the $\Lambda$CDM cosmography $j(z)=1$, comes out as
\begin{equation}\label{epsilon_values}
    \epsilon =
    \begin{cases}
        0.03305& \qquad\qquad \text{(Evolutionary model-I)}
        \\
        0.035& \qquad\qquad \text{(Evolutionary model-II)}
    \end{cases}\,.
\end{equation}

Some words are in order regarding the imposition of the condition $\Omega_{m0}=\frac{2}{3}(1+q_0)$. This condition holds for the $\Lambda$CDM \emph{model} itself, but not for any other model that is cosmographically equivalent to it (e.g. the dark fluid model discussed at the end of Section \ref{sec:kin_vs_dyn}). If one considers the limit of exact $\Lambda$CDM-like evolution $\epsilon=0$, which can be obtained by setting $j_0=1$ (see Eqs. \eqref{const_redef-I} and \eqref{const_redef-II}), one can see that it does not automatically make $w_{\rm DE}(z)|_{\epsilon\to0}=-1$. This is again a confirmation of the fact that a $\Lambda$CDM-like cosmic evolution does not necessarily imply the $\Lambda$CDM \emph{model} itself. However, if one imposes the condition $\Omega_{m0}=\frac{2}{3}(1+q_0)$, then one gets $w_{\rm DE}(z)|_{\epsilon\to0}=-1$. Only then can we safely conclude that the smallness parameter $\epsilon$ not only characterize a deviation from the $\Lambda$CDM-like cosmic evolution, but also a deviation from the $\Lambda$CDM \emph{model} as well.

In fact, it is seen from Eqs. \eqref{wDE_cosmology-I} and \eqref{wDE_cosmology-II} that the imposition of the condition $\Omega_{m0}=\frac{2}{3}(1+q_0)$ makes $w_{\rm DE}(z=0)=-1$ for both the models even without the $\epsilon\to0$ limit. Therefore, the underlying models characterizing the almost $\Lambda$CDM evolution actually coincide with the exact $\Lambda$CDM model \emph{today}.

\begin{figure}[H]
    \centering    
    \begin{subfigure}{0.49\linewidth}
        \includegraphics[width=\linewidth]{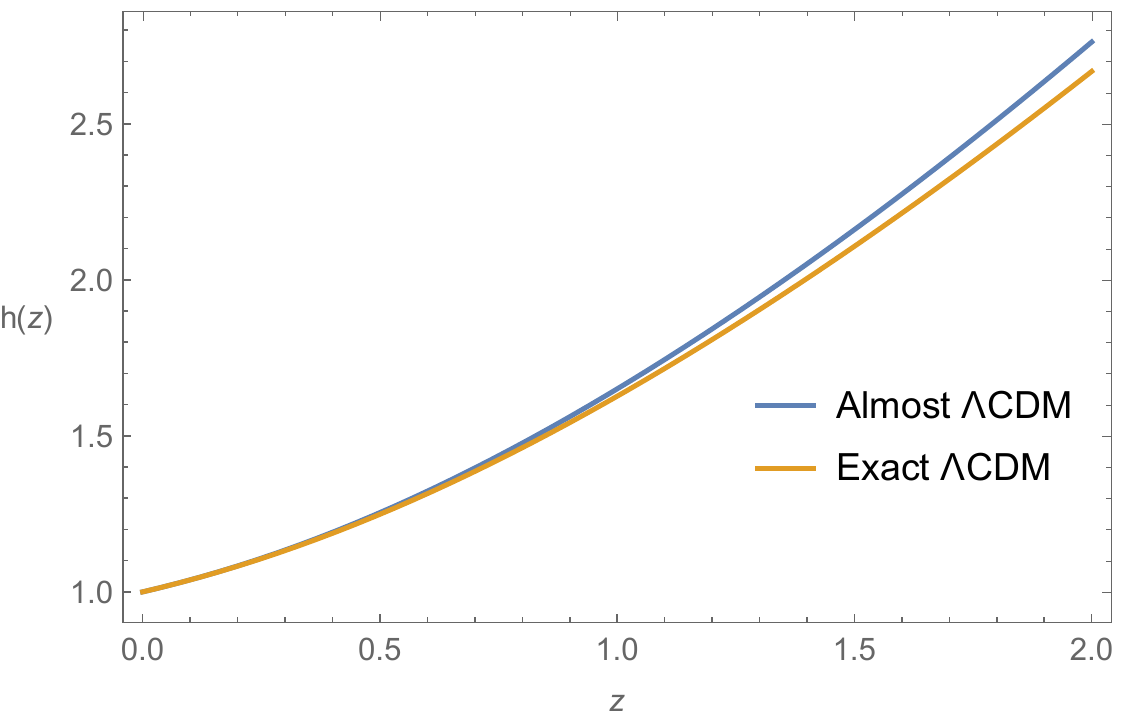}
        \caption{}
        \label{fig:h1}
    \end{subfigure}
    \hfill
    \begin{subfigure}{0.49\linewidth}
        \includegraphics[width=\linewidth]{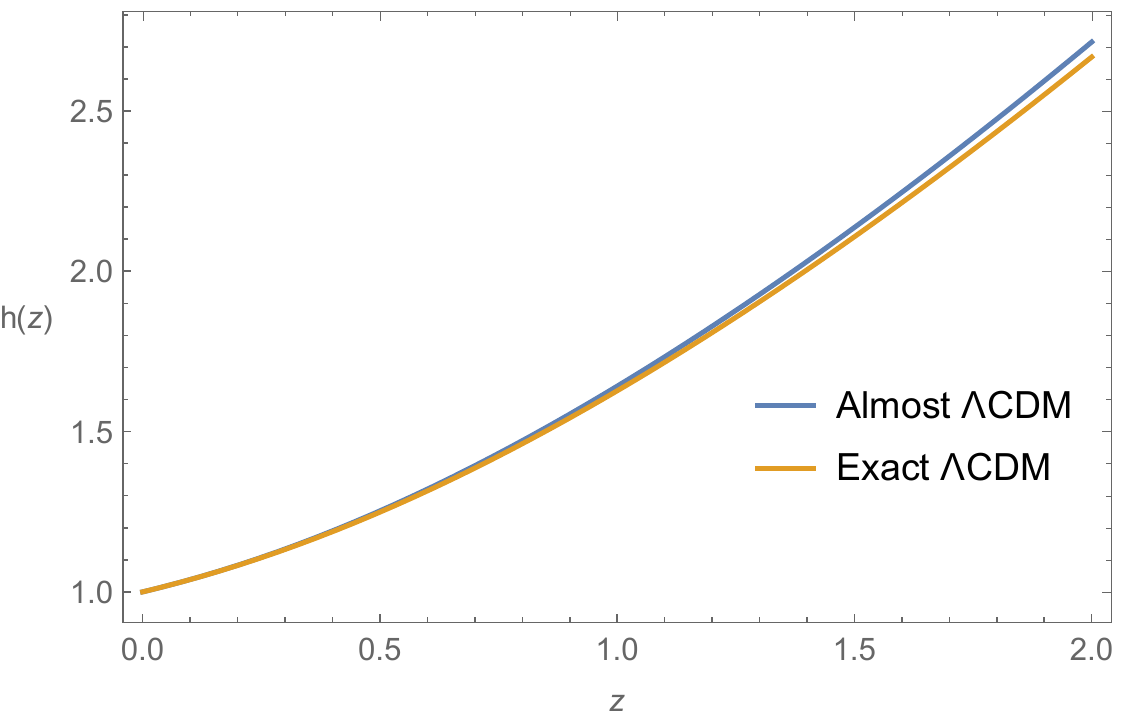}
        \caption{}
        \label{fig:h2}
    \end{subfigure}
    \begin{subfigure}{0.49\textwidth}
        \includegraphics[width=\linewidth]{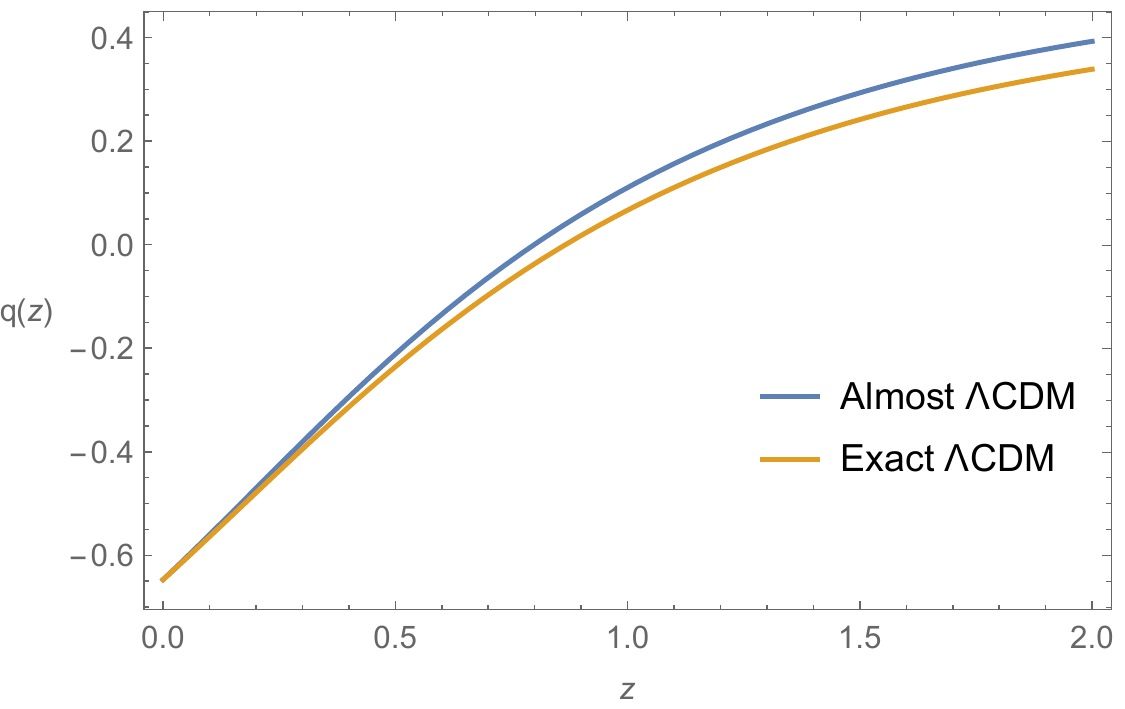}
        \caption{}
        \label{fig:q1}
    \end{subfigure}
    \hfill
    \begin{subfigure}{0.49\textwidth}
        \centering
        \includegraphics[width=\linewidth]{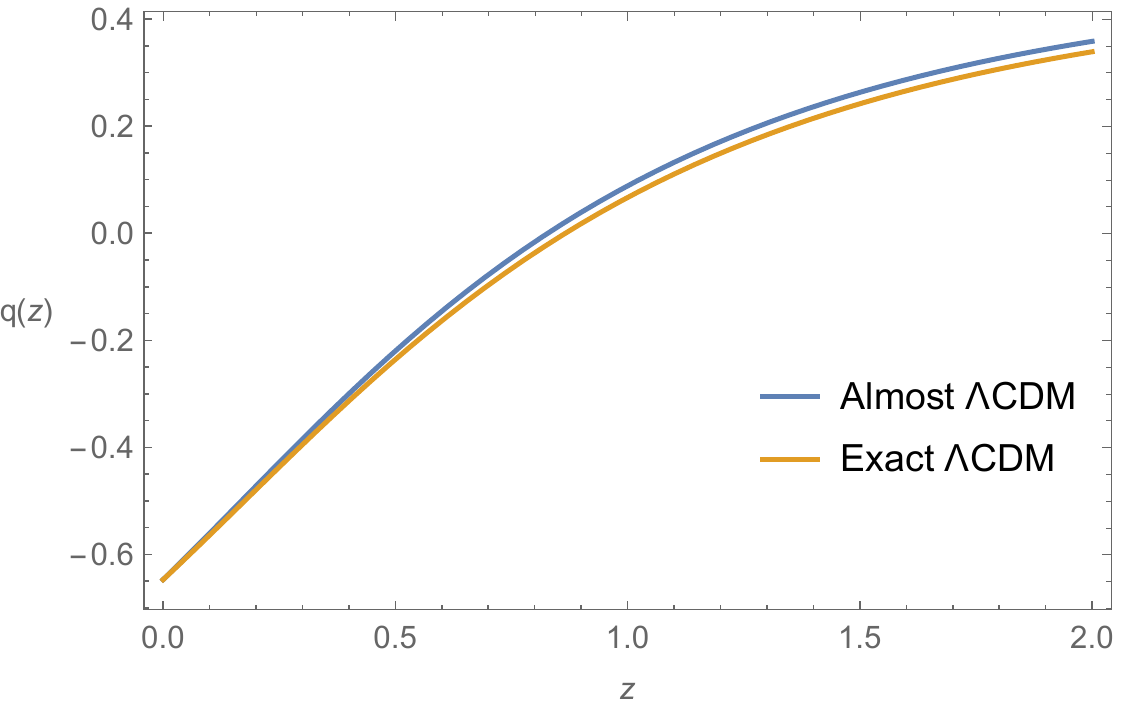}
        \caption{}
        \label{fig:q2}
    \end{subfigure}
    \begin{subfigure}{0.49\textwidth}
        \includegraphics[width=\linewidth]{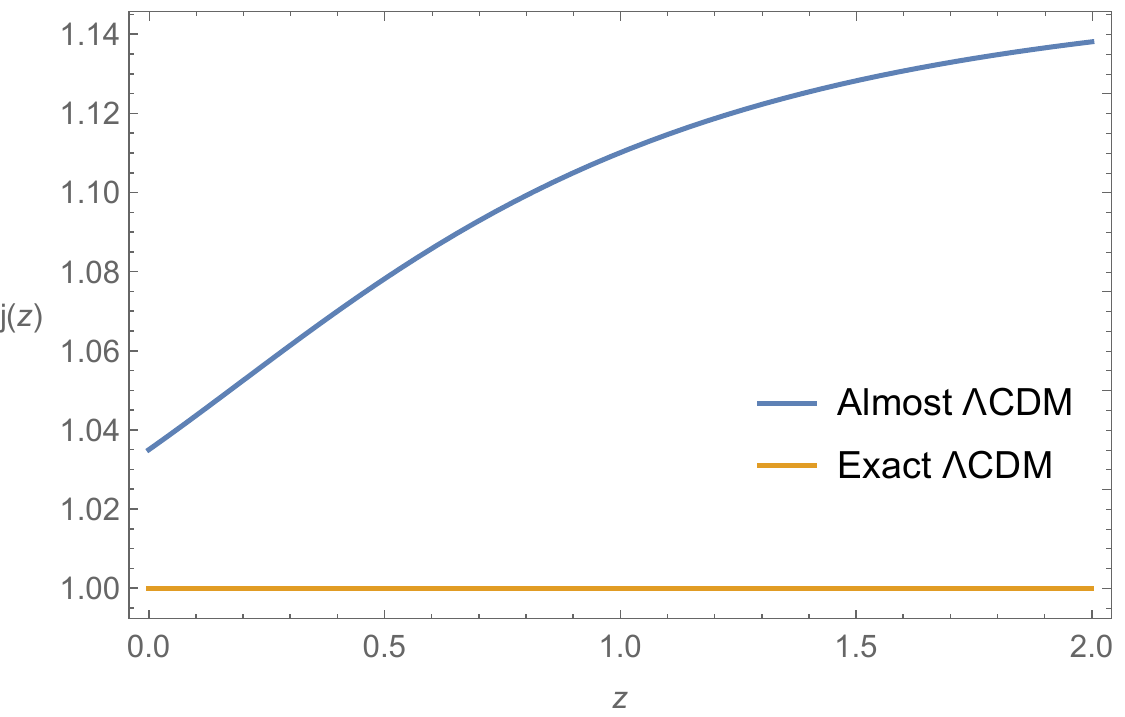}
        \caption{}
        \label{fig:j1}
    \end{subfigure}
    \hfill
    \begin{subfigure}{0.49\textwidth}
        \centering
        \includegraphics[width=\linewidth]{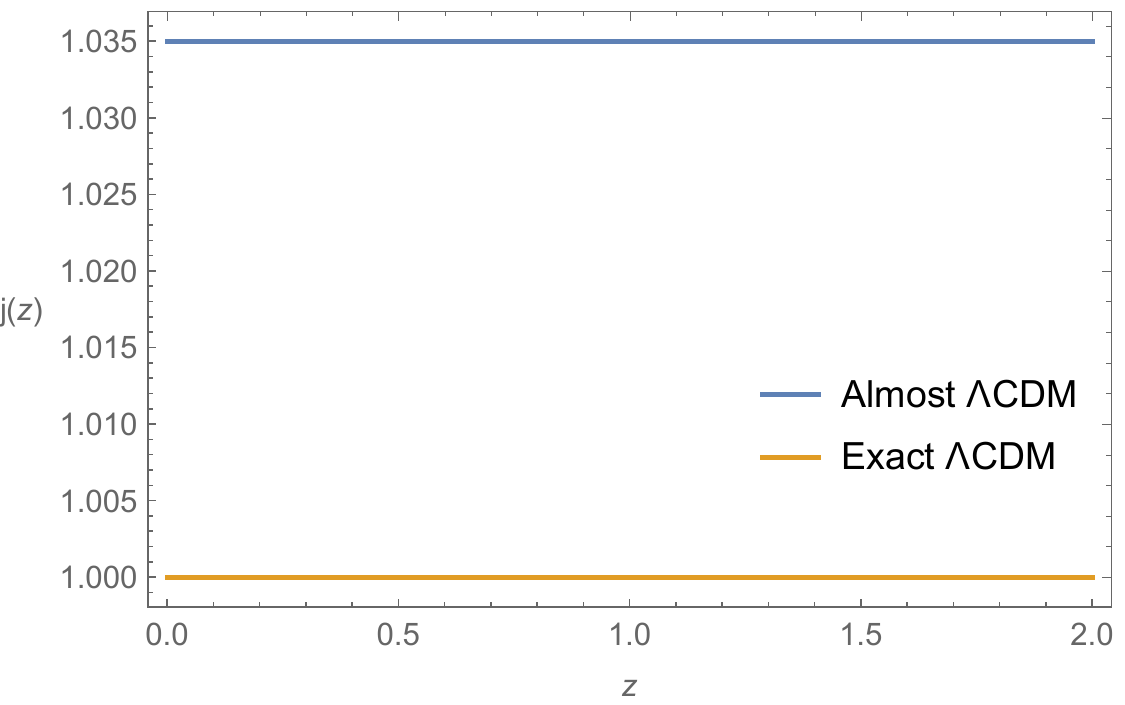}
        \caption{}
        \label{fig:j2}
    \end{subfigure}
    \begin{subfigure}{0.49\textwidth}
        \centering
        \includegraphics[width=\linewidth]{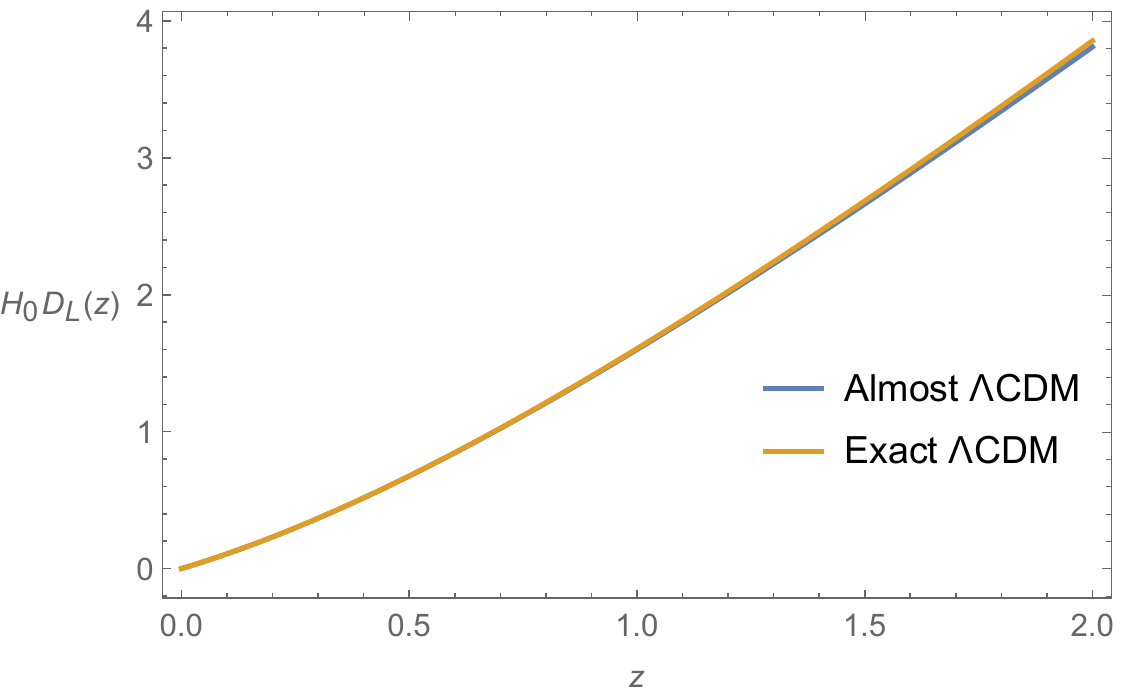}
        \caption{}
        \label{fig:dL1}
    \end{subfigure}
    \hfill
    \begin{subfigure}{0.49\textwidth}
        \centering
        \includegraphics[width=\linewidth]{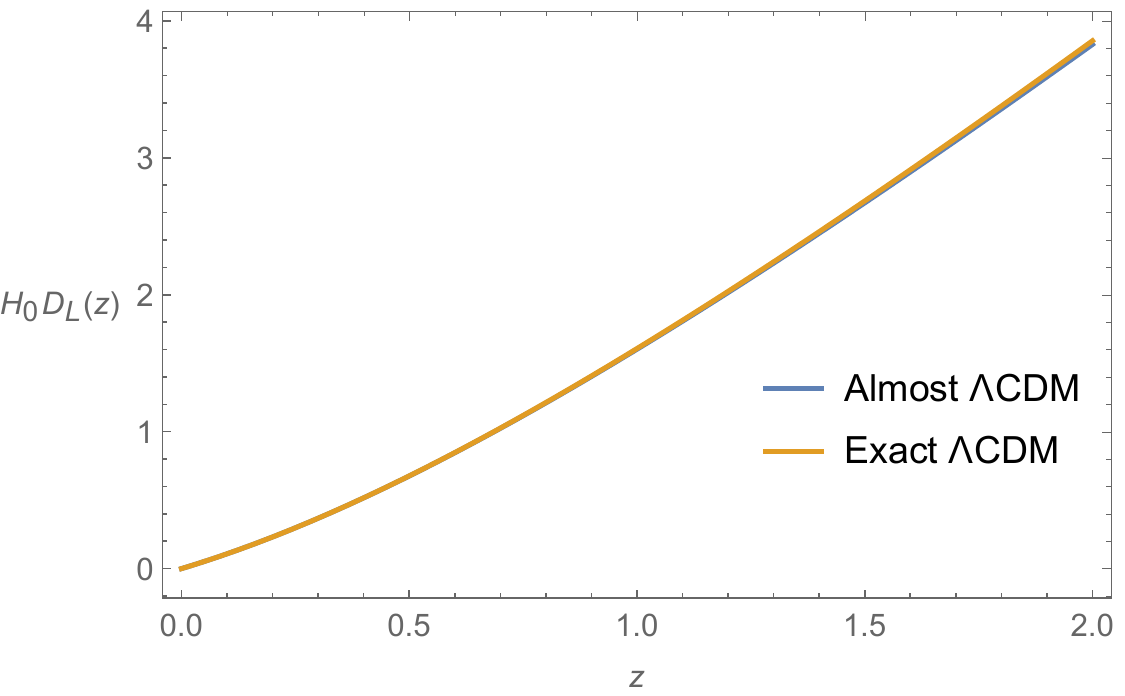}
        \caption{}
        \label{fig:dL2}
    \end{subfigure}
     \caption{Evolution of different kinematic quantities for the almost $\Lambda$CDM evolutionary model-I (left panels) and model-II (right panel). In the last row, we also show the plots of the quantity $H_0 D_L$ for the two evolutionary models, where $D_L=(1+z)\int_0^z \frac{dz}{H(z)}$ is the luminosity distance.}
     \label{fig:kinematic}
\end{figure}

\begin{figure}[H]
    \centering
    \begin{subfigure}{0.49\textwidth}
        \centering
        \includegraphics[width=\linewidth]{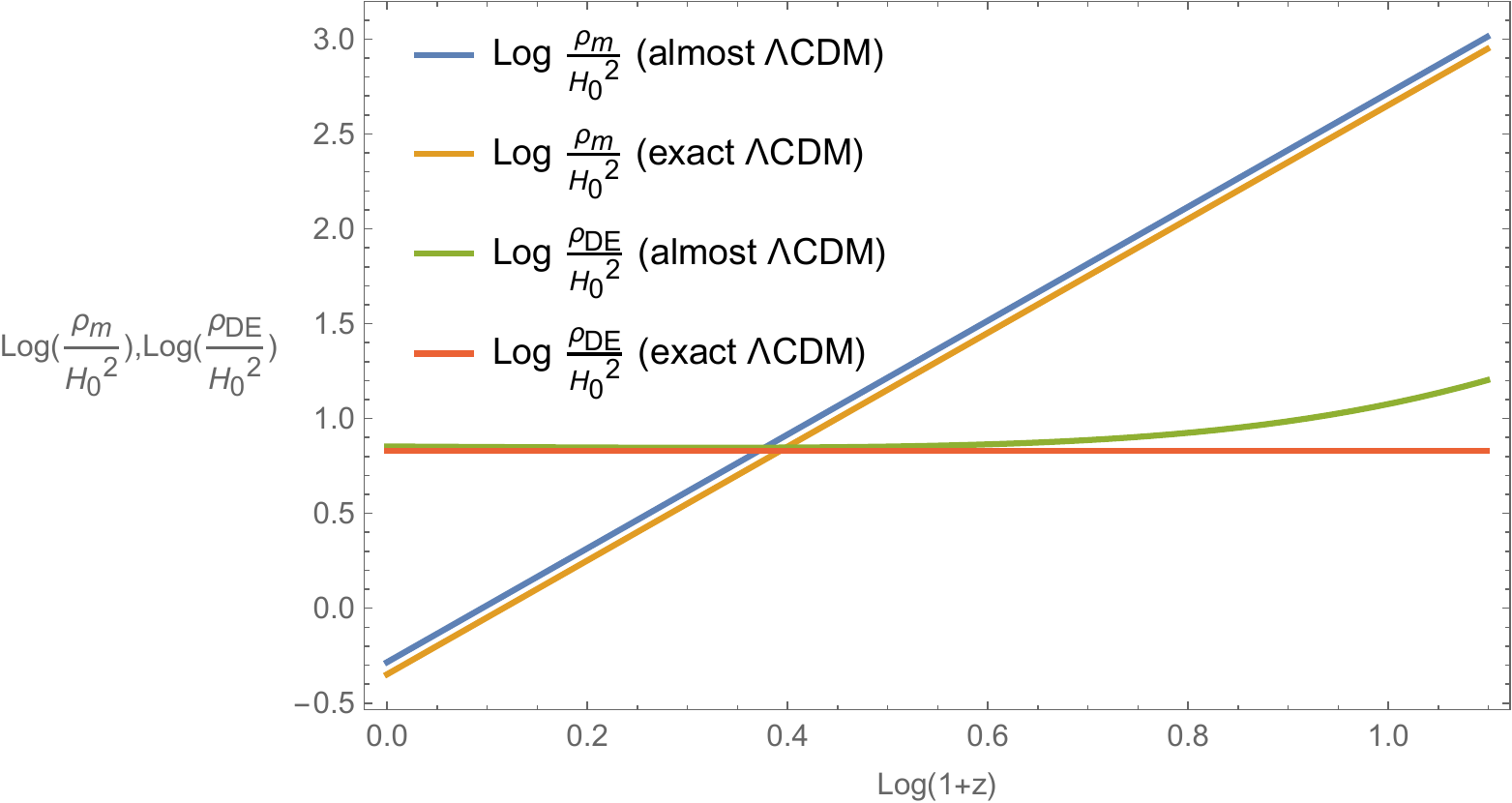}
        \caption{}
        \label{fig:rho1}
    \end{subfigure}
    \hfill
    \begin{subfigure}{0.49\textwidth}
        \centering
        \includegraphics[width=\linewidth]{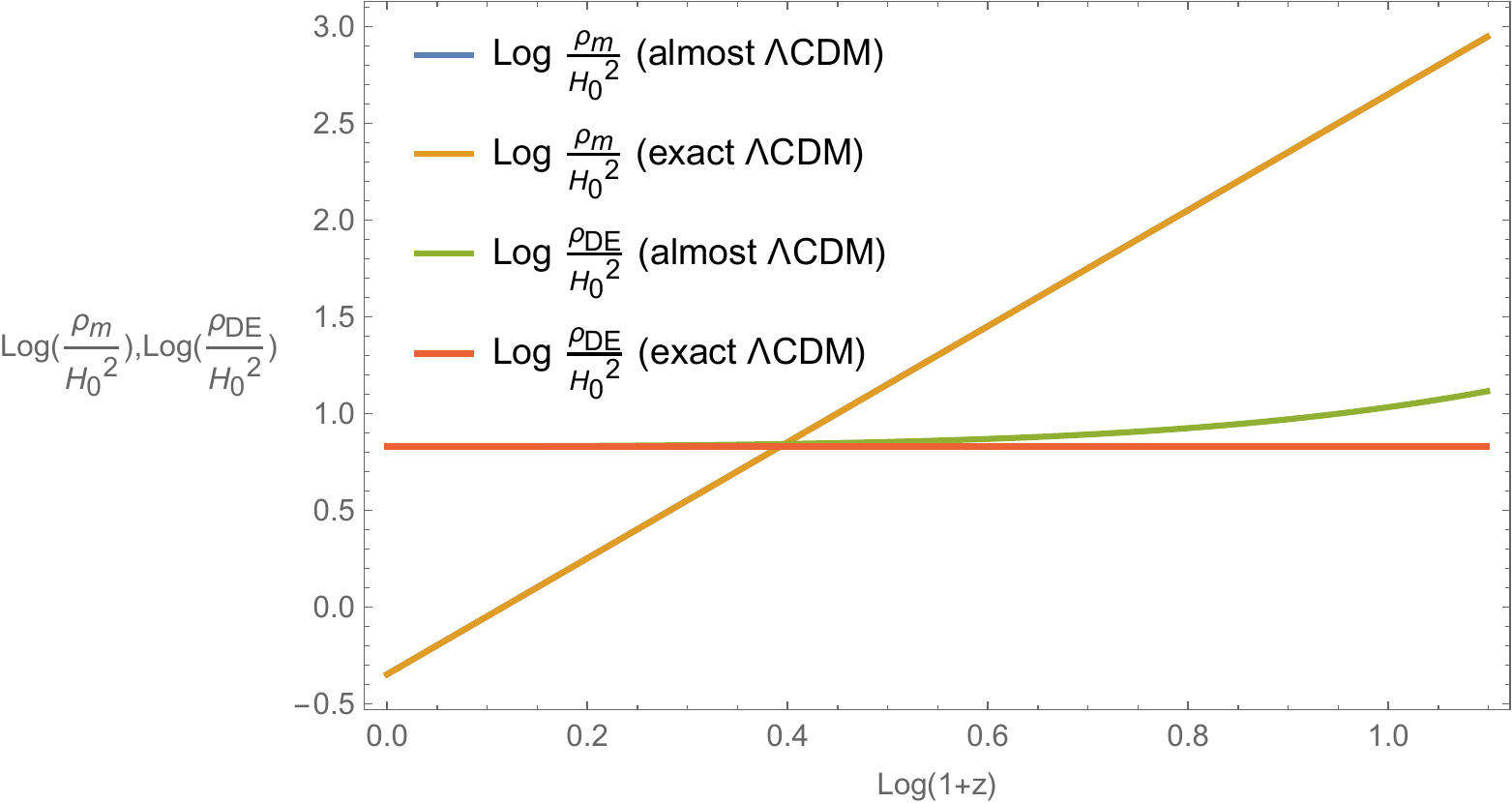}
        \caption{}
        \label{fig:rho2}
    \end{subfigure}
    \begin{subfigure}{0.49\textwidth}
        \centering
        \includegraphics[width=\linewidth]{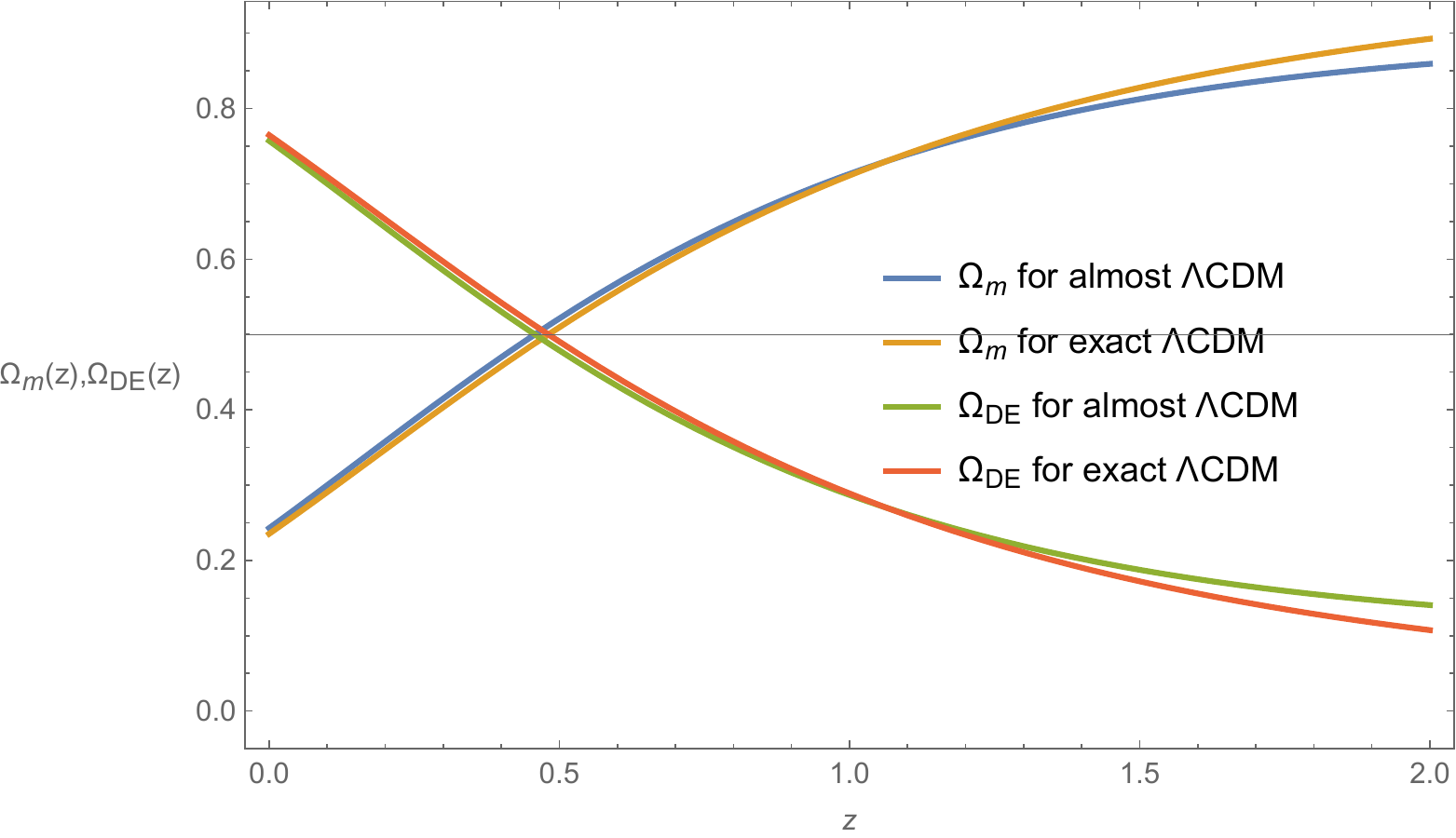}
        \caption{}
        \label{fig:Omega1}
    \end{subfigure}
    \hfill
    \begin{subfigure}{0.49\textwidth}
        \centering
        \includegraphics[width=\linewidth]{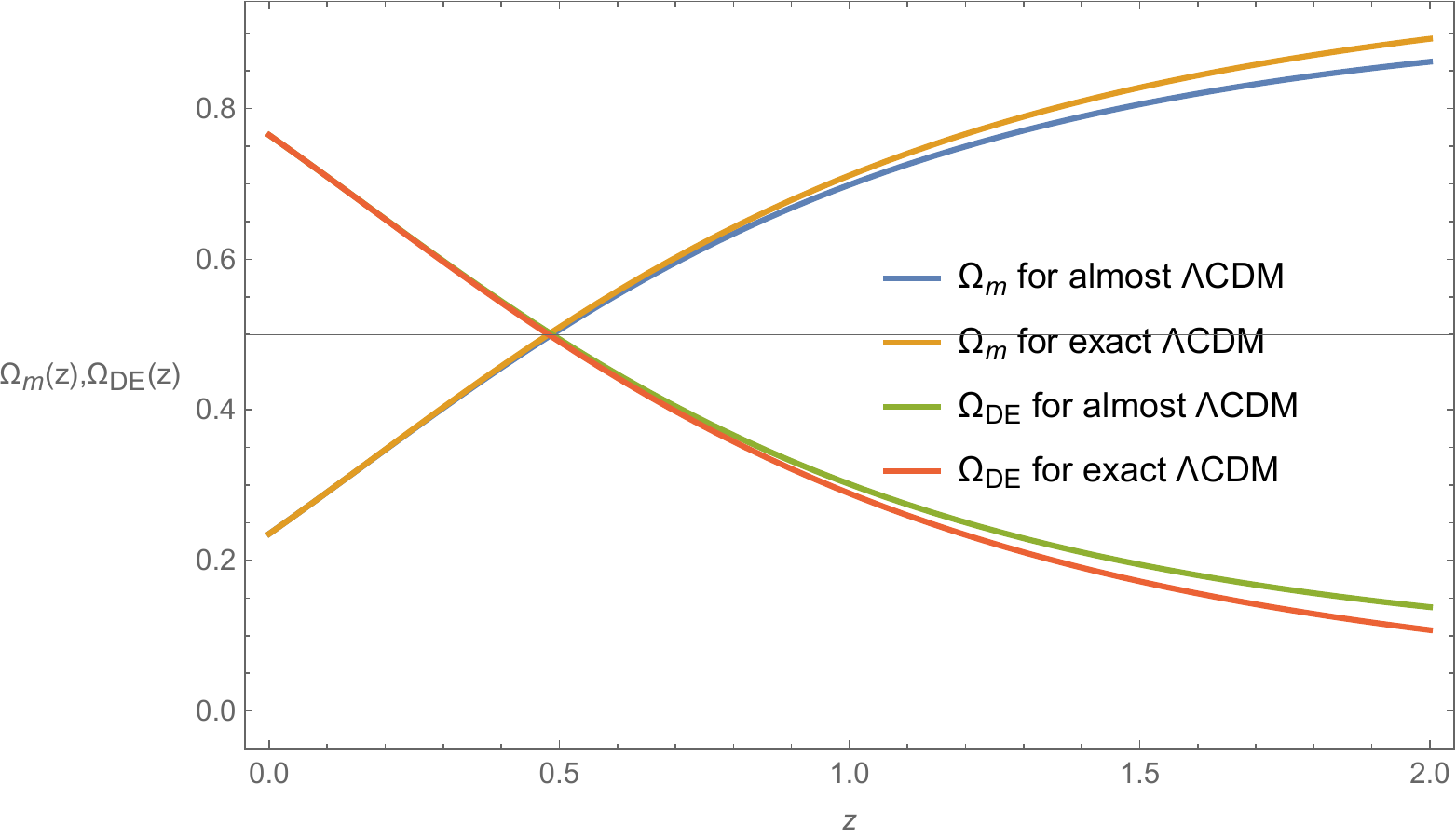}
        \caption{}
        \label{fig:Omega2}
    \end{subfigure}
    \caption{Evolution of different densities and the density abundance parameters for the almost $\Lambda$CDM evolutionary model-I (left panels) and model-II (right panel).}
    \label{fig:density}
\end{figure}

\begin{figure}[H]
    \centering
    \begin{subfigure}{0.49\textwidth}
        \includegraphics[width=\linewidth]{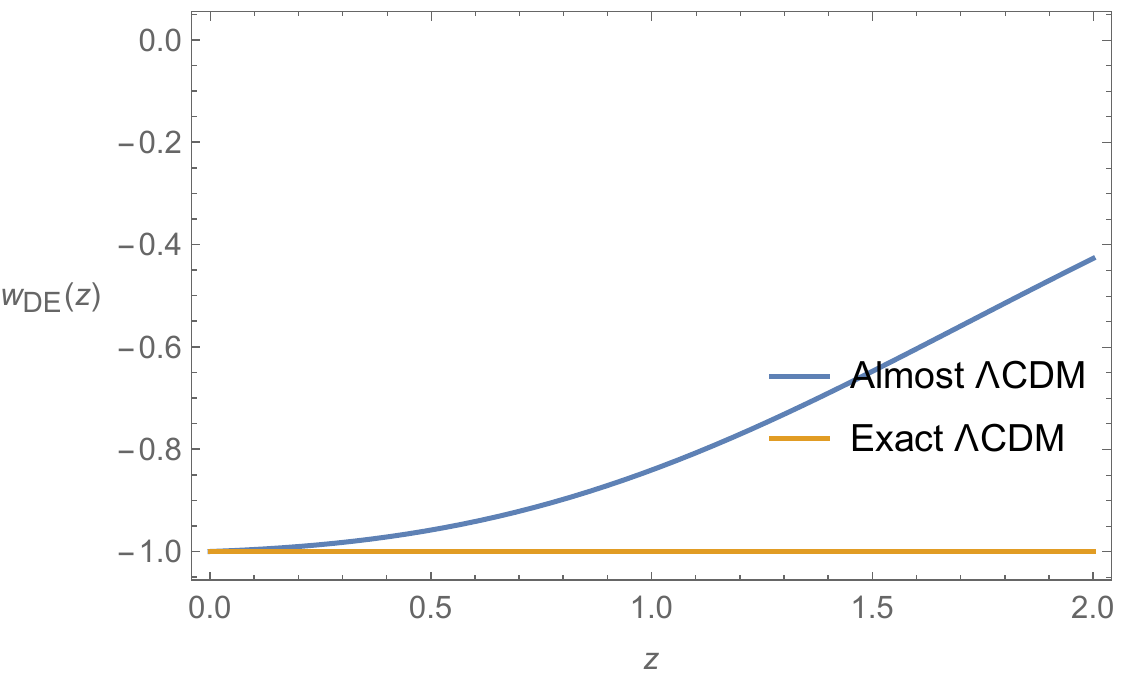}
        \caption{}
        \label{fig:wde1}
    \end{subfigure}
    \hfill
    \begin{subfigure}{0.49\textwidth}
        \centering
        \includegraphics[width=\linewidth]{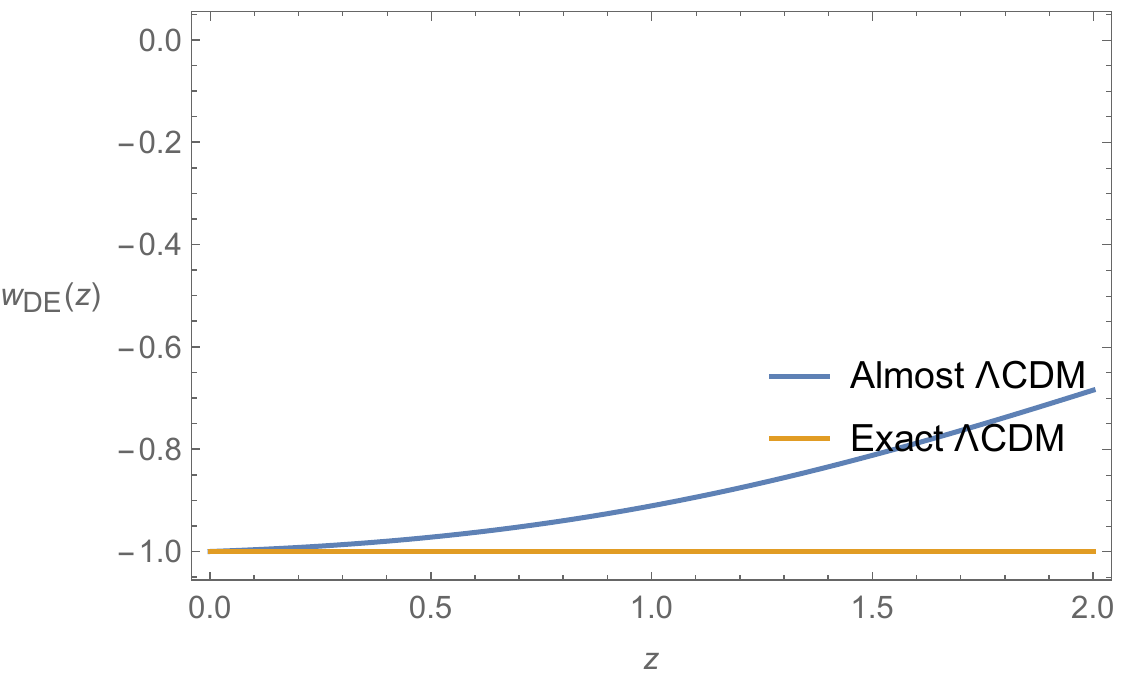}
        \caption{}
        \label{fig:wde2}
    \end{subfigure}
    \caption{Evolution of the dark energy equation of state parameter for the almost $\Lambda$CDM evolutionary model-I (left panels) and model-II (right panel).}
    \label{fig:wde}
\end{figure}

\begin{figure}[H]
    \centering
    \begin{subfigure}{0.49\textwidth}
        \centering
        \includegraphics[width=\linewidth]{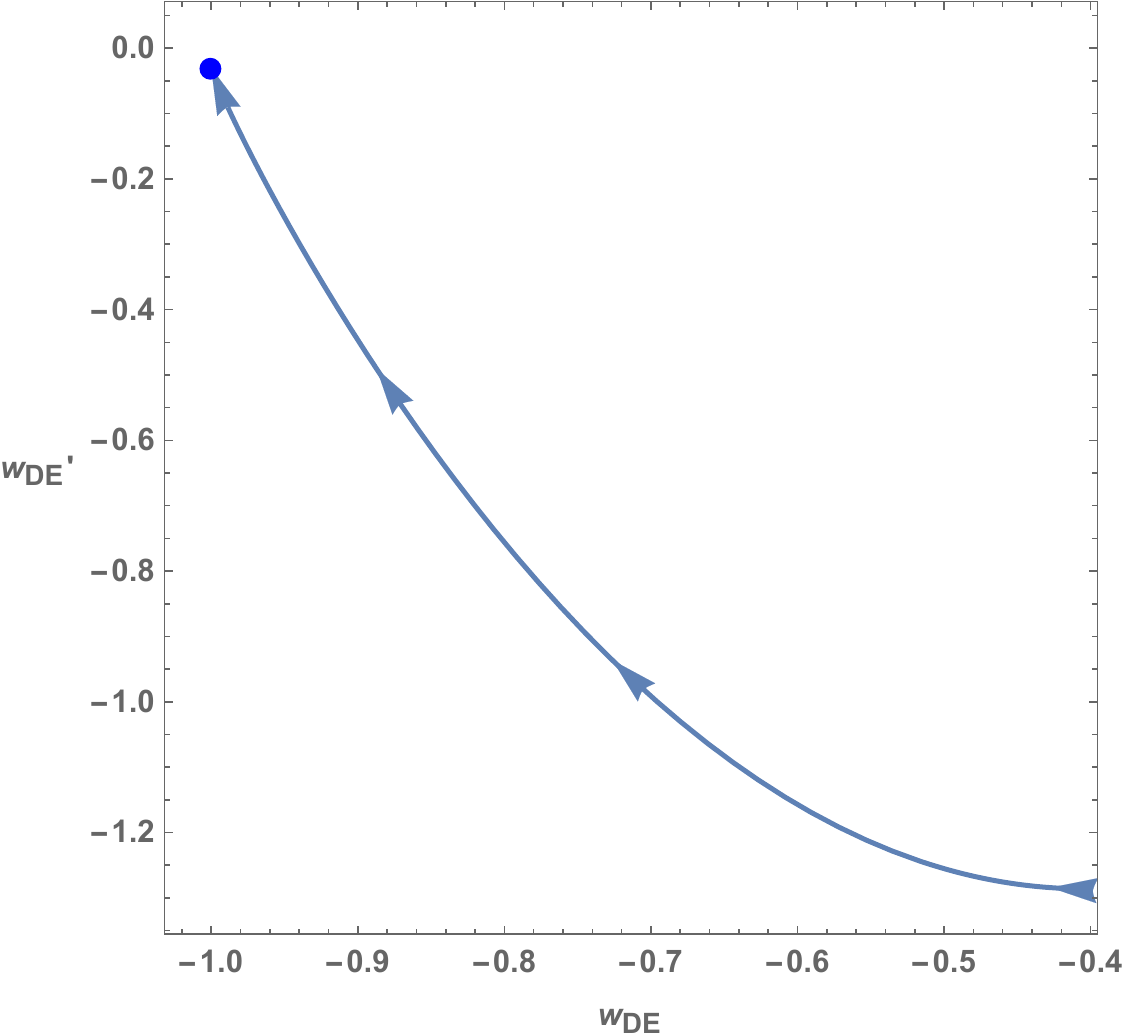}
        \caption{}
        \label{fig:ww'1}
    \end{subfigure}
    \hfill
    \begin{subfigure}{0.49\textwidth}
        \centering
        \includegraphics[width=\linewidth]{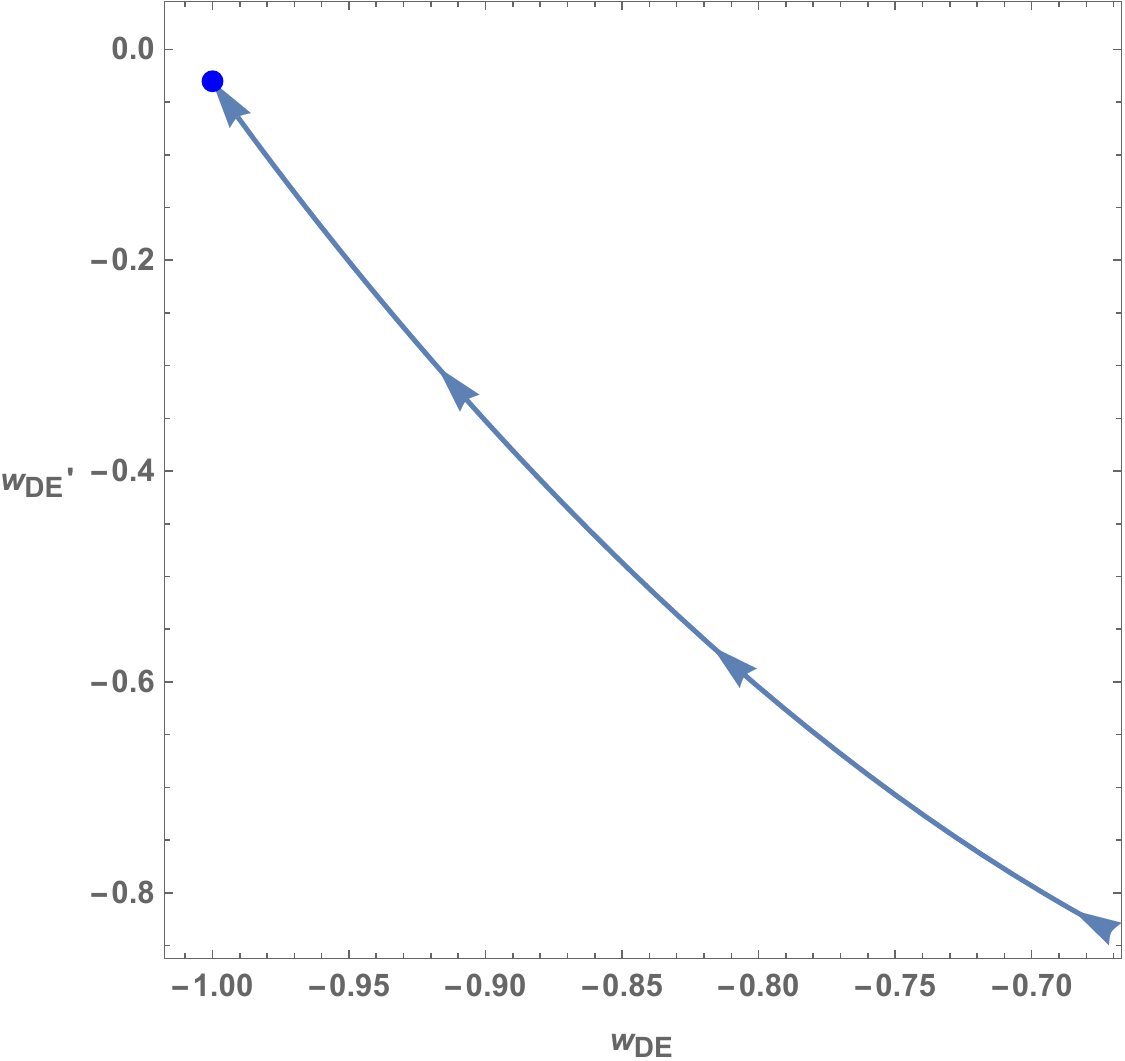}
        \caption{}
        \label{fig:ww'2}
    \end{subfigure}
    \caption{Evolution of the dark energy equation of state in the $w-w'$ plane for the almost $\Lambda$CDM evolutionary model-I (left panels) and model-II (right panel). Here $w_{\rm DE}'\equiv=\frac{dw_{\rm DE}}{d\log(a)}=-(1+z)\frac{dw_{\rm DE}}{dz}$. For both the models, $\{w_{\rm DE}(0),w'_{\rm DE}(0)\}=\{-1,-0.0305144\}$, i.e. they coincide with the $\Lambda$CDM model $w_{\rm DE}=-1$ at $z=0$.}
    \label{fig:ww'}
\end{figure}

The time evolutions of various kinematic cosmological quantities is shown in Fig. \ref{fig:kinematic}. In all the plots, the evolution of quantities corresponding to the exact $\Lambda$CDM-like evolution is obtained by setting $j_0=1$, which makes $\epsilon=0$.
The time evolutions of the densities and the density abundance parameters are shown in Fig.\ref{fig:density}. 
The evolution of the dark energy equation of states of the underlying model for both the almost $\Lambda$CDM-like evolutions is shown in Fig.\ref{fig:wde}
The evolution of $w_{\rm DE}(z)$ is portrayed in the $w-w'$ plane ($w'=\frac{dw}{d\ln(a)}$) in Fig. \ref{fig:ww'}. Interestingly, it can be seen that the expressions of $w_{\rm DE}(z=0)$ and $w_{\rm DE}'(z=0)$ in terms of $q_0,j_0,\Omega_{m0}$ are the same for both the almost $\Lambda$CDM evolutionary models, as can be seen from Eqs. \eqref{w0wa-I} and \eqref{w0wa-II}. For the chosen parameter values \eqref{param_values}, one has
\begin{equation}
    \left\lbrace w_{\rm DE},\frac{{\rm d} w_{\rm DE}}{{\rm d}\ln(a)} \right\rbrace_{a=a_0=1} = \left\lbrace w_{\rm DE},-(1+z)\frac{{\rm d}w_{\rm DE}}{{\rm d}z} \right\rbrace_{z=0} = \{-1,-0.0305144\}\,.
\end{equation}

The interesting noteworthy point here is the evolution of the dark energy equation of state $w_{\rm DE}$. By imposing $\Omega_{m0}=\frac{2}{3}(1+q_0)$ we have set things up such that the exact $\Lambda$CDM-like evolution corresponds to the exact $\Lambda$CDM model as well. This is apparent from Fig. \ref{fig:wde}. One would naively expect that small deviation from the $\Lambda$CDM-like evolution would correspond to a small deviation in the $\Lambda$CDM model as well, i.e. small deviation in the behaviour of $w_{\rm DE}$. It turns out that this is not the case. Even though all the \emph{kinematic} quantities like the Hubble, deceleration, jerk parameter and the luminosity distance show small deviations, model-dependent quantities that depend on the underlying dynamics, e.g. energy densities, density abundance parameters, and the dark energy equation of state show significant deviation. In other words, one can conclude that kinematic closeness to the $\Lambda$CDM model does not necessarily guarantee dynamical closeness to the $\Lambda$CDM model. 

The above conclusion has serious ramifications when trying to compare models against cosmographic data. Even though a cosmological model may appear very close to the $\Lambda$CDM model according to the luminosity distance redshift data, the model itself can be quite different from the $\Lambda$CDM model. Put in another way, even though the $\Lambda$CDM model may offer a very good fit with the luminosity distance redshift data, one cannot use this fact to confidently state that the underlying model \emph{is} actually the $\Lambda$CDM model, or something very close to it.

\begin{figure}[H]
    \centering
    \begin{subfigure}{0.49\textwidth}
        \includegraphics[width=\linewidth]{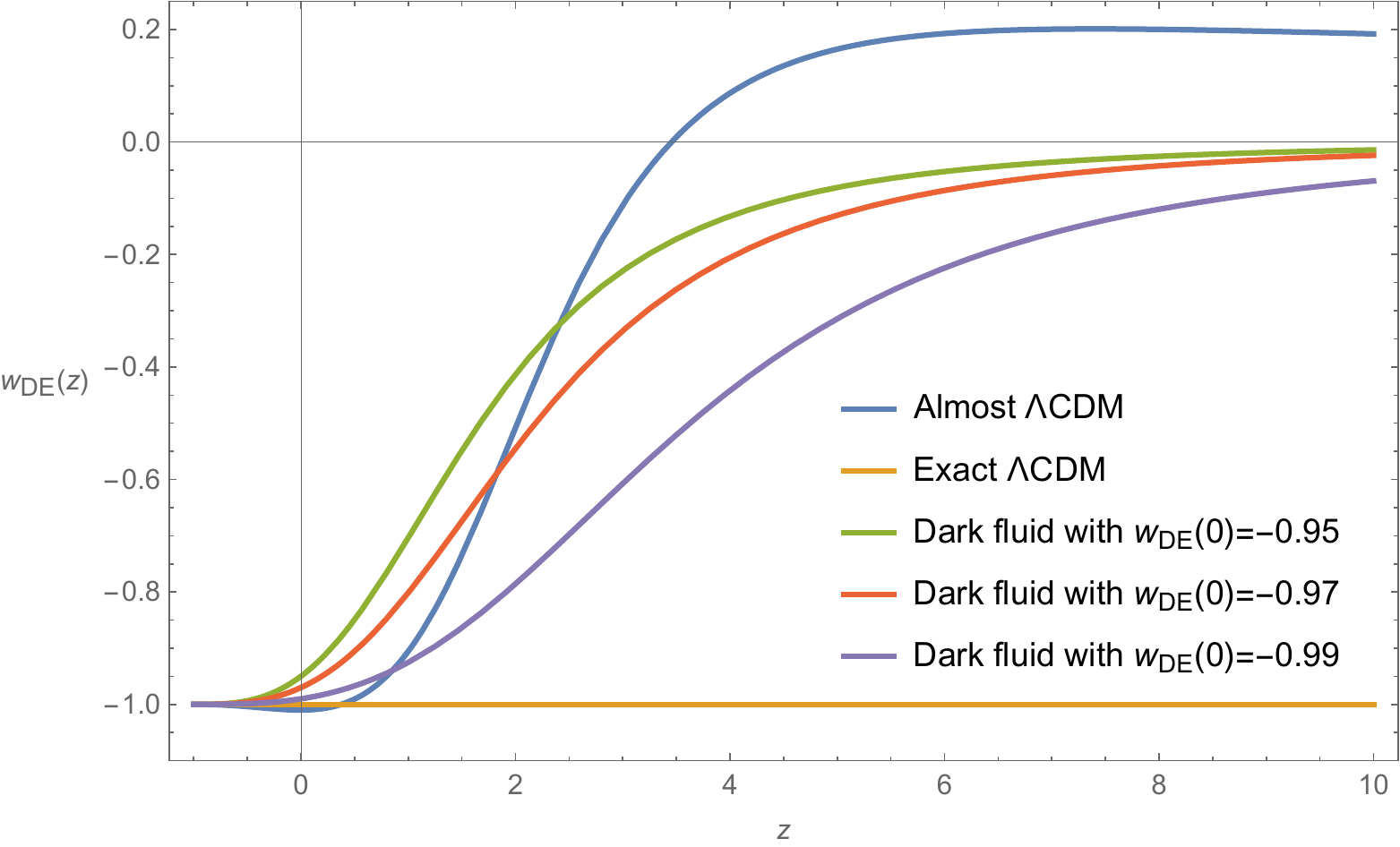}
        \caption{}
        \label{fig:wde1_df}
    \end{subfigure}
    \hfill
    \begin{subfigure}{0.49\textwidth}
        \centering
        \includegraphics[width=\linewidth]{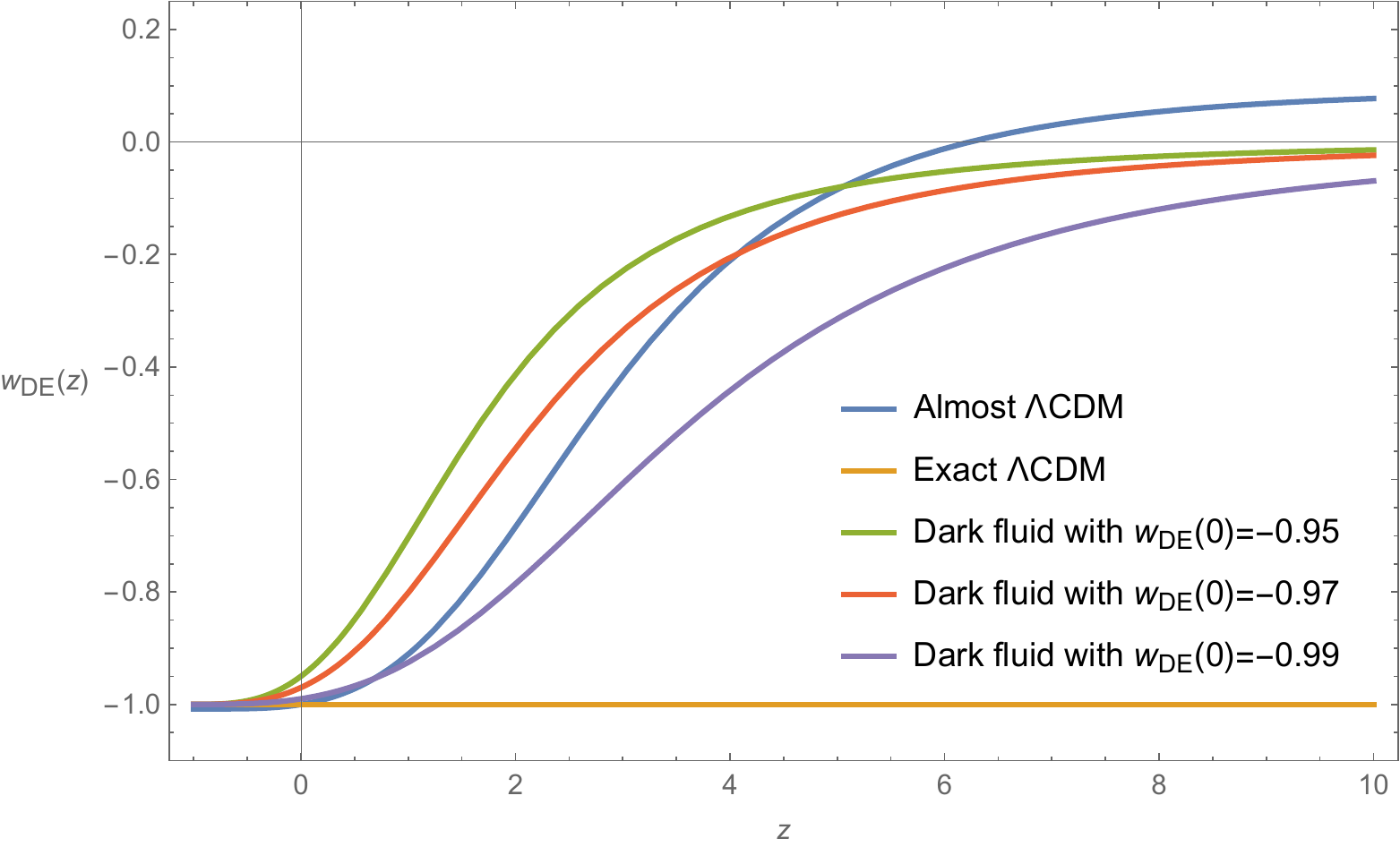}
        \caption{}
        \label{fig:wde2_df}
    \end{subfigure}
    \caption{Evolution of the dark energy equation of state parameter for the almost $\Lambda$CDM evolutionary model-I (left panels) and model-II (right panel).}
    \label{fig:wde_df}
\end{figure}

In Section \ref{sec:kin_vs_dyn}, we have mentioned that the condition $j(z)=1$ generically can imply not only the $\Lambda$CDM model but also the unified dark fluid model. It is interesting to note that, even though we have set things up such that the exact $\Lambda$CDM evolution $j(z)=1$ does actually correspond to the exact $\Lambda$CDM model (i.e. imposition of the condition $\Omega_{m0}=\frac{2}{3}(1+q_0)$), it turns out that almost $\Lambda$CDM evolutions rather correspond to almost dark fluid model. This is apparent when one compares the behaviors of $w_{\rm DE}(z)$ corresponding to the almost $\Lambda$CDM evolutionary models over a larger redshift range (see Fig. \ref{fig:wde_df}); the respective $w_{\rm DE}(z)$s asymptote to a value close to (but not exactly) zero at high redshift and asymptotes to $-1$ at low redshift. 
This, in a sense, implies that the $\Lambda$CDM model $w_{\rm DE}=-1$ is actually not a robust one among all the possibilities offered by the cosmographic condition $j(z)=1$. Given that the condition $j(z)=1$ is not an exact but only an approximated one as obtained from the luminosity distance vs redshift data, a unified dark fluid model is a far more likely and robust theoretical option for a model.

\section{Compatibility with the DESI results}\label{sec:DESI}

We have shown that, both the almost $\Lambda$CDM evolutionary models considered here (Eqs. \eqref{almostlcdm_1} and \eqref{almostlcdm_2}) correspond generically to a dynamical dark energy model (Eqs. \eqref{wDE_cosmology-I} and \eqref{wDE_cosmology-II}). In this section, we intend to critically examine the two evolutionary models in light of the latest DESI data, which strongly hints towards the possibility of a dynamical dark energy scenario. For this purpose, we use \cite[Table V]{DESI:2025zgx}, which uses the Chevallier-Polarski-Linder (CPL) parametrization to parameterize the dark energy equation of state
\begin{equation}\label{CPL}
    w_{\rm DE} = w_0 + w_a(1-a) = w_0 + w_a\left(\frac{z}{1+z}\right)\,.
\end{equation}
Since we have been neglecting any contribution from the global spatial curvature, the pertinent sector from \cite[Table V]{DESI:2025zgx} for us is the $w_0w_aCDM$ model. It can be calculated that
\begin{equation}
    \{w_0,w_a\} = \lbrace w_{\rm DE}(0),w_{\rm DE}'(0)\rbrace\,.
\end{equation}

In the last section, while obtaining the numerical solutions and the corresponding plots, we have taken the values of the parameters $\{q_0,j_0\}$ from \cite[Table 5]{Mukherjee:2020ytg}, which were obtained in a model-independent manner by combining the CC+BAO+Pantheon+CMB data and employing a Gaussian process. In this section, we will not specify the values of $\{q_0,j_0\}$ \emph{a-priori}, but rather obtain their values, or a bound on them, on the requirement that the underlying dynamical dark energy models that give rise to the almost $\Lambda$CDM evolutions considered, are consistent with the DESI results \cite{DESI:2025zgx}.

Firstly, let us set $\Omega_{m0}=\frac{2}{3}(1+q_0)$, which makes $w_0 = w_{\rm DE}(0) = -1$ for both the models. Considering different combinations of datasets, \cite[Table V]{DESI:2025zgx} provides the value of the parameter $w_0$ within the range
\begin{equation}\label{w0_DESI}
    -1.23 < w_0 < -0.42\,
\end{equation}
at $68\%$ confidence level\footnote{We have ignored the error bars.}, which contains the value $w_0=-1$. One can now set $\Omega_{m0}=\frac{2}{3}(1+q_0)$ at Eqs.\eqref{w0wa-I} and \eqref{w0wa-II} to obtain
\begin{equation}
    w_a = w_{\rm DE}'(0) = - \frac{j_0 - 1}{q_0 - 1/2}\,.
\end{equation}
An attentive reader can recognize the above expression to be equal to $-3s_0$, $s$ being one of the statefinder parameters used in statefinder diagnostic \cite{Sahni:2002fz,Alam:2003sc}. This is expected, since the statefinder parameter $s$ is related to the dark energy equation of state as
\begin{eqnarray}
    s = 1 + w_{\rm DE} - \frac{1}{3 w_{\rm DE}}\frac{{\rm d} w_{\rm DE}}{{\rm d} \log(a)} = 1 + w_{\rm DE} + \frac{1}{3 w_{\rm DE}}(1+z)\frac{{\rm d} w_{\rm DE}}{{\rm d} z}\,.
\end{eqnarray}
If $w_{\rm DE}(z=0)=-1$, then $\frac{{\rm d} w_{\rm DE}}{{\rm d} z}\vert_{z=0} = -3s_0$. Considering different combinations of datasets, \cite[Table V]{DESI:2025zgx} provides the value of the parameter $w_a$ within the range
\begin{equation}\label{wa_DESI}
    -1.75 < w_a < -0.17\,,
\end{equation}
which translates into a bound on the values of $\{q_0,j_0\}$
\begin{equation}\label{q0j0_bound}
  0.17 < \frac{j_0 - 1}{q_0 - 1/2} < 1.75\,. 
\end{equation}

The DESI collaboration paper \cite{DESI:2025zgx} does not put any model-independent constraint on the present-day values of the deceleration and jerk parameters $\{q_0,j_0\}$. For the almost $\Lambda$CDM evolutionary models considered in this paper to be consistent with the latest DESI data, they must satisfy the bound \eqref{q0j0_bound}. One can note that the values of $\{q_0,j_0\}$ chosen in Eq.\eqref{q0j0_values}, although providing a cosmography very close to the $\Lambda$CDM model, do \emph{not} actually satisfy the bound \eqref{q0j0_bound}. The DESI data therefore, hints towards a cosmology that is even cosmographically very different from that of the $\Lambda$CDM model. A recent work on cosmography after DESI's second data release \cite{Rodrigues:2025tfg} appears to reach a similar conclusion.

Next, let us not even fix $\Omega_{m0}$ \emph{a-priori}, but take the values of $\{\Omega_{m0},w_0,w_a\}$ from \cite[Table V]{DESI:2025zgx}. Since $\{w_0,w_a\} = \lbrace w_{\rm DE}(0),w_{\rm DE}'(0)\rbrace$, this allows us to calculate the values of $\{q_0,j_0\}$ using \eqref{w0wa-I} and \eqref{w0wa-II}. The results are shown in Table \ref{tab:q0j0_table}.
\begin{table}[H]
    \centering
    \begin{tabular}{c|c|c|c|c|c|c|c|}
       Dataset  & $\Omega_m$ & $\omega_0$ & $\omega_a$ & $q_0$ & $j_0$ & $\epsilon$ (Model I) & $\epsilon$ (Model II)\\\hline
       DESI+Pantheon+ & $0.298$ & $-0.888$ & $-0.17$ & $-0.435$ & $0.507$ & $-0.29$& $-0.49$\\
       DESI+Union3 & $0.328$ & $-0.70$ & $-0.99$ & $-0.206$ & $-0.633$ & $-0.685$ & $-1.633$\\
       DESI+DESY5 & $0.319$ & $-0.781$ & $-0.72$ & $-0.298$ & $-0.26$ & $-0.598$ & $-1.26$\\
       DESI+$(\theta_*,\omega_b,\omega_c)_{CMB}$ & $0.353$ & $-0.43$ & $-1.72$ & $0.083$ & $-1.383$ & $-0.734$ & $-2.383$\\
       DESI+CMB (no lensing)& $0.352$ & $-0.43$ & $-1.70$ & $0.082$ & $-1.367$ & $-0.729$ & $-2.367$\\
       DESI+CMB& $0.353$ & $-0.42$ & $-1.75$ & $0.0924$ & $-1.408$ & $-0.735$ & $-2.408$\\
       DESI+CMB+Pantheon+ & $0.3114$ & $-0.838$ & $-0.62$ & $-0.366$ & $-0.061$ & $-0.557$ & $-1.061$\\
       DESI+CMB+Union3 & $0.3275$ & $-0.667$ & $-1.09$ & $-0.173$ & $-0.772$ & $-0.714$ &$-1.772$\\
       DESI+CMB+DESY5 & $0.3191$ & $-0.752$ & $-0.86$ & $-0.268$ & $-0.45$ & $-0.66$ & $-1.45$
    \end{tabular}
    \caption{The numerical values of the deviation parameter $\epsilon$ that quantifies the deviation of the cosmography from the cosmography of the $\Lambda$CDM model ($j(z)=1$) are listed for various combinations of data sets used in the latest DESI data release (see \cite[Table V]{DESI:2025zgx}; error bars not taken into consideration). One can notice that the absolute values of the deviation parameter obtained from the DESI data are orders of magnitude bigger than what was obtained in the previous section (see Eq. \eqref{epsilon_values}), pushing the resulting cosmography far from that of the $\Lambda$CDM. This emphasizes the fact that the DESI data \emph{probably} hints towards a cosmological model whose cosmography is quite different from that of the $\Lambda$CDM model.}
    \label{tab:q0j0_table}
\end{table}

From the calculated values of $\{q_0,j_0\}$, one can also calculate the corresponding values of the deviation parameter $\epsilon$ from Eqs. \eqref{const_redef-I} and \eqref{const_redef-II} for both the evolutionary models. These values are listed in the last two columns of Table \ref{tab:q0j0_table}. As one can notice, the absolute values of the deviation parameter obtained are orders of magnitude bigger compared to what was obtained in the previous section (see Eq. \eqref{epsilon_values}), which pushes the resulting cosmography far from that of the $\Lambda$CDM. This reaffirms the possibility that the latest DESI data \emph{possibly} hints towards a cosmological model whose cosmography is quite different from that of the $\Lambda$CDM model.

\begin{figure}[H]
    \centering    
    \begin{subfigure}{0.49\linewidth}
        \includegraphics[width=\linewidth]{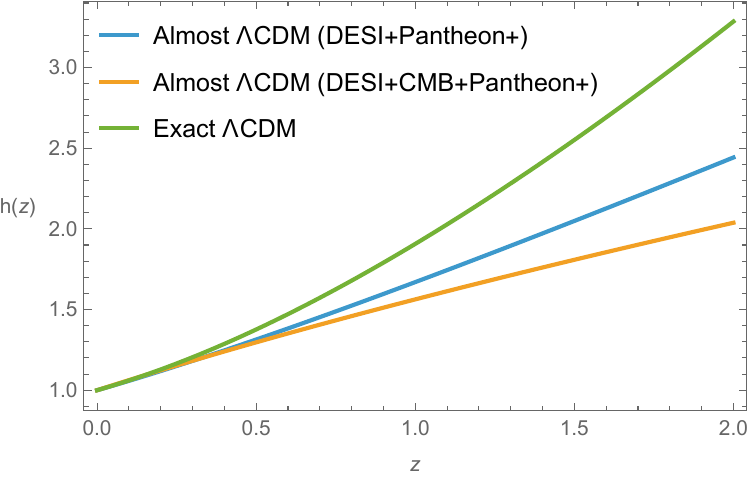}
        \caption{}
        \label{fig:h1_DESI}
    \end{subfigure}
    \hfill
    \begin{subfigure}{0.49\linewidth}
        \includegraphics[width=\linewidth]{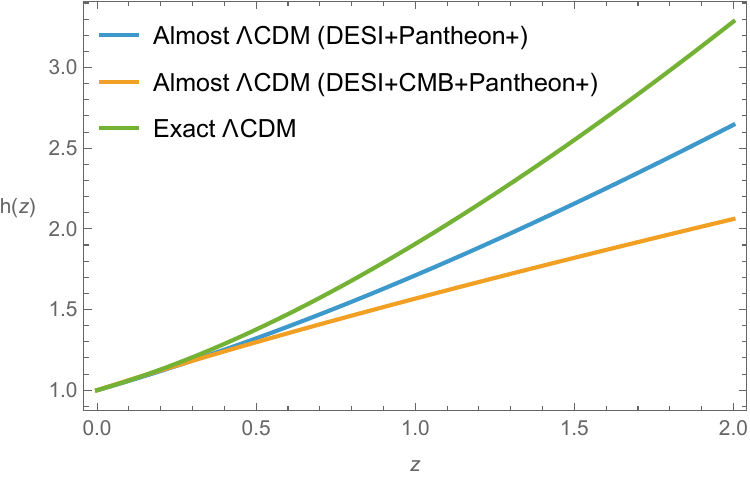}
        \caption{}
        \label{fig:h2_DESI}
    \end{subfigure}
    \begin{subfigure}{0.49\textwidth}
        \includegraphics[width=\linewidth]{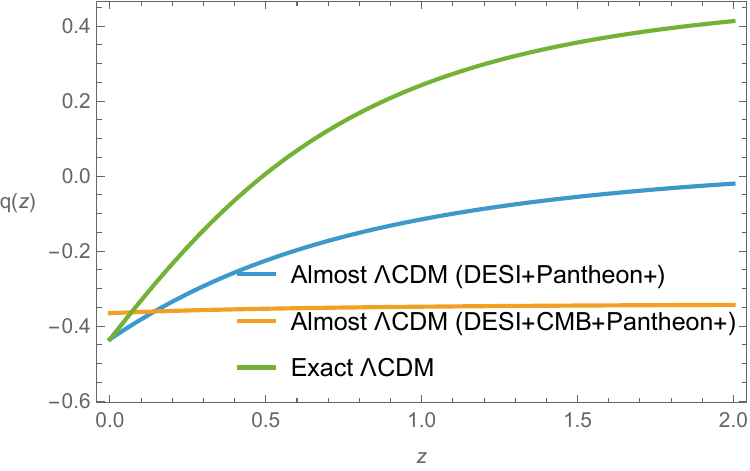}
        \caption{}
        \label{fig:q1_DESI}
    \end{subfigure}
    \hfill
    \begin{subfigure}{0.49\textwidth}
        \centering
        \includegraphics[width=\linewidth]{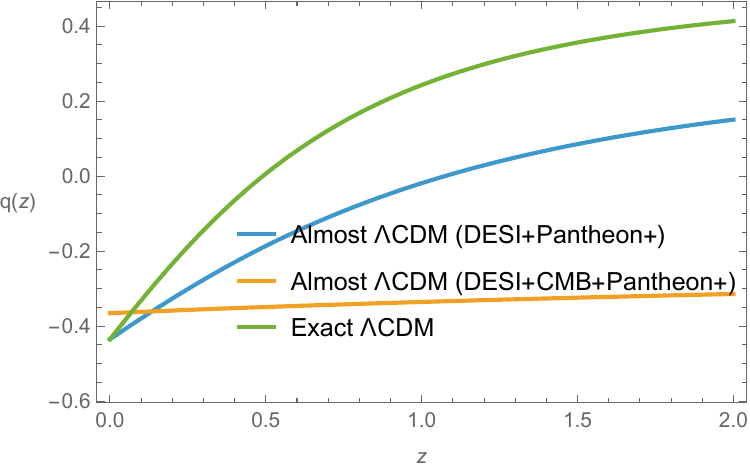}
        \caption{}
        \label{fig:q2_DESI}
    \end{subfigure}
    \begin{subfigure}{0.49\textwidth}
        \includegraphics[width=\linewidth]{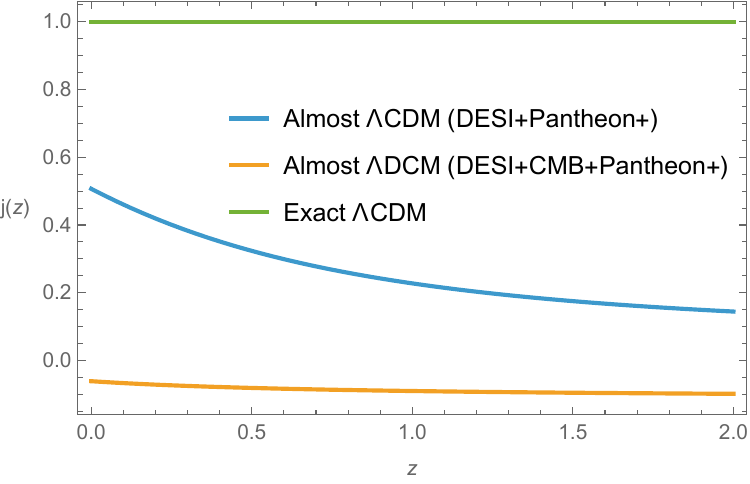}
        \caption{}
        \label{fig:j1_DESI}
    \end{subfigure}
    \hfill
    \begin{subfigure}{0.49\textwidth}
        \centering
        \includegraphics[width=\linewidth]{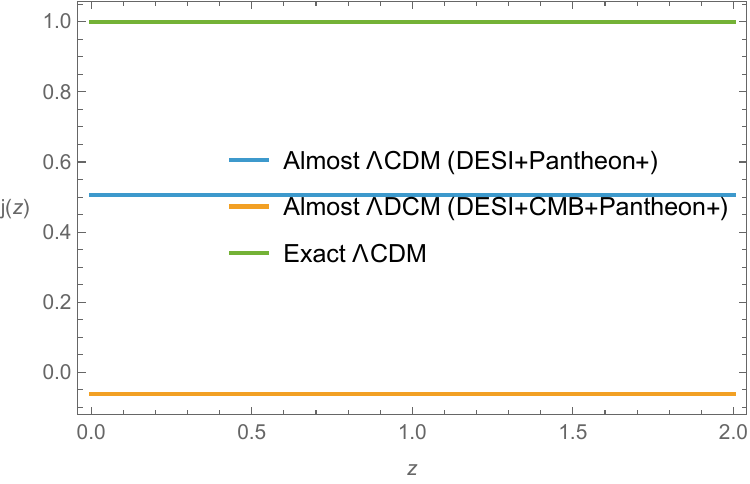}
        \caption{}
        \label{fig:j2_DESI}
    \end{subfigure}
    \begin{subfigure}{0.49\textwidth}
        \centering
        \includegraphics[width=\linewidth]{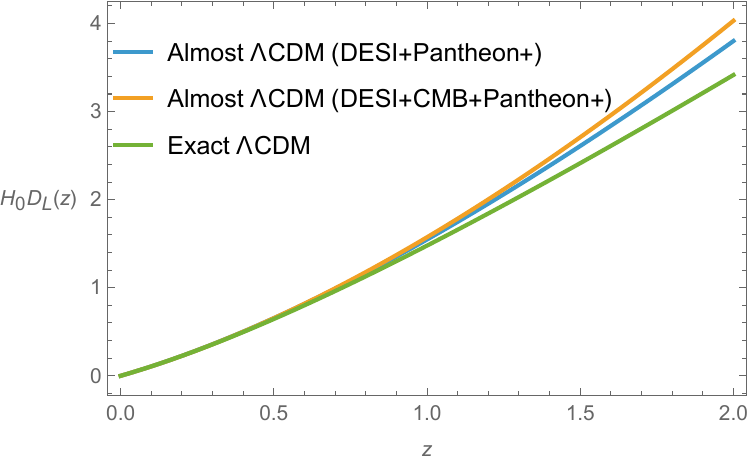}
        \caption{}
        \label{fig:DL1_DESI}
    \end{subfigure}
    \hfill
    \begin{subfigure}{0.49\textwidth}
        \centering
        \includegraphics[width=\linewidth]{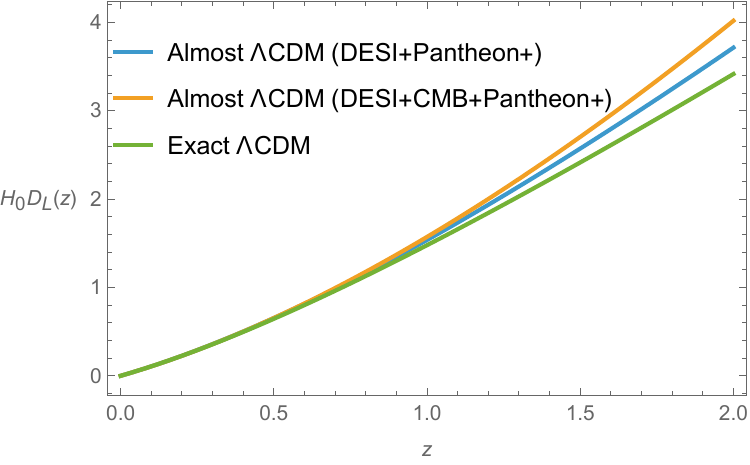}
        \caption{}
        \label{fig:DL2_DESI}
    \end{subfigure}
     \caption{Evolution of different kinematic quantities for the almost $\Lambda$CDM evolutionary model-I (left panels) and model-II (right panel). Two datasets were chosen from Table \ref{tab:q0j0_table}: DESI+Pantheon+ (in blue) and DESI+CMB+Pantheon (in orange), and the plots correspond to the respective $\epsilon$ values.}
     \label{fig:kinematic_DESI}
\end{figure}

Two data sets were chosen, namely the DESI+Pantheon+ and the DESI+CMB+Pantheon+ dataset, in order to investigate this difference in cosmography between the $\Lambda$CDM model and the latest DESI data. The time evolution of various kinematic cosmological quantities using these datasets are plotted in Fig. \ref{fig:kinematic_DESI}. Once again the evolution of quantities corresponding to the exact $\Lambda$CDM-like evolution were obtained by setting $j_0=1$, making $\epsilon=0$.

\begin{figure}[H]
    \centering
    \begin{subfigure}{0.49\textwidth}
        \includegraphics[width=\linewidth]{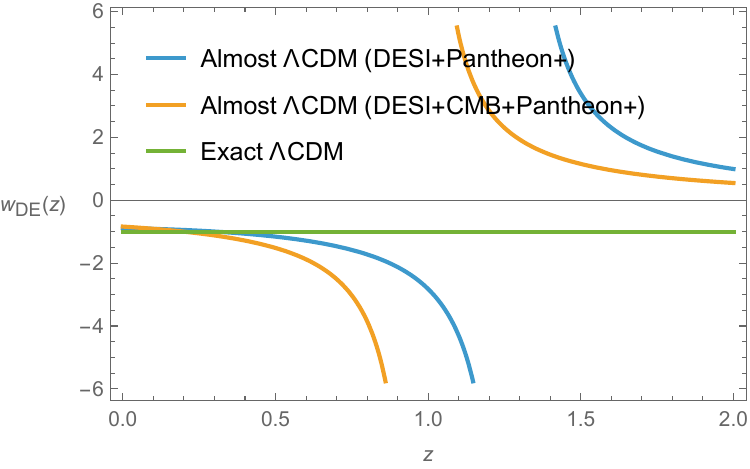}
        \caption{}
        \label{fig:wde1_DESI}
    \end{subfigure}
    \hfill
    \begin{subfigure}{0.49\textwidth}
        \centering
        \includegraphics[width=\linewidth]{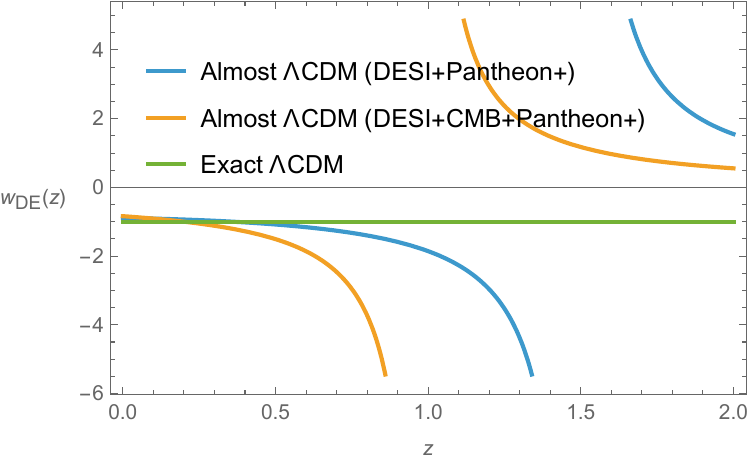}
        \caption{}
        \label{fig:wde2_DESI}
    \end{subfigure}
    \caption{Evolution of the dark energy equation of state parameter for the almost $\Lambda$CDM evolutionary model-I (left panels) and model-II (right panel). Two datasets were chosen from Table \ref{tab:q0j0_table}: DESI+Pantheon+ (in blue) and DESI+CMB+Pantheon (in orange), and the plots correspond to the respective $\epsilon$ values. Model-I has a behaviour similar to $w_{\rm DE}$ in Tsujikawa's $f(R)$ model \cite{Tsujikawa:2007xu}.}
    \label{fig:wde_DESI}
\end{figure}
In Fig. \ref{fig:wde_DESI} it is clear that for both models, using the DESI+CMB+Pantheon+ dataset results in a dark energy equation of state parameter that becomes phantom quickly while the DESI+Pantheon+ dataset follows the exact $\Lambda$CDM-like evolution more closely.

\section{Conclusion}\label{sec:conclusion}
In this investigation we explored the robustness of the equivalence between the kinematical and dynamical descriptions of the spatially flat $\Lambda$CDM model. We began by reaffirming that the well-known cosmographic condition $j=1$ is in fact degenerate between the $\Lambda$CDM model and the unified dark-fluid model. In fact, $j=1$ indeed corresponds to the $\Lambda$CDM model only if we take $w_{\rm DE}(0)$ in \eqref{eq:DDE} to be exactly equal to $-1$. In all other cases $j=1$ does not correspond to the $\Lambda$CDM model. In other words, the $\Lambda$CDM model is of measure zero in the solution space characterized by $j=1$. 

This observation led us to explore what happens if one considers models that are cosmographically close to $\Lambda$CDM. We achieved this by constructing two explicit models, where the deviation away from $\Lambda$CDM is determined by the parameter $\epsilon$. The first can be thought of as a homogeneous perturbation of the $\Lambda$CDM scale factor, while the second represents a small deviation away from the $j=1$ condition. 

We found that even if an exact $\Lambda$CDM-like cosmic evolution corresponds to the exact $\Lambda$CDM model, an almost $\Lambda$CDM-like cosmic evolution does not imply an almost $\Lambda$CDM model. This has implications for cosmological model building because cosmographic data always give $j(z)\approx1$ and never $j(z)=1$ exactly. It follows that an almost dark fluid model is more strongly favored by cosmographic data.

We also used the latest DESI data to constrain the value of our deviation parameter $\epsilon$, giving values that result in a cosmography far from what is obtained for $\Lambda$CDM. This conclusion is in agreement with cosmographic studies performed after DESI's second data release \cite{Rodrigues:2025tfg}. Although only two models were investigated in this analysis, we expect that other small deviations would result in similar results. The examples considered here suffice to showcase the non-equivalence of kinematic and dynamic specifications of a model.

\section*{Acknowledgements}

SC is supported by the Second Century Fund (C2F), Chulalongkorn University, Thailand. PKSD acknowldges support from the First Rand Bank (South Africa).
AdlCD acknow\-ledges support from BG20/00236 action (MCINU, Spain), NRF Grant CSUR23\-042798041, CSIC Grant COOPB23096, Project SA097P24 funded by Junta
de Castilla y Le\'on (Spain) and Grants PID2024-158938NB-I00 and PID2021-122938NB-I00 funded by MCIN / AEI (Spain) and by {\it ERDF A way of making Europe}.

\bibliography{references}
\bibliographystyle{unsrt}

\end{document}